\documentclass[aps,prb,twocolumn, amsmath, amssymb, superscriptaddress]{revtex4-2}
\usepackage{graphicx}
\usepackage{bm}
\usepackage{amsmath}
\usepackage{float}
\usepackage[FIGTOPCAP]{subfigure}
\usepackage[framemethod=tikz]{mdframed}
\usepackage{bbold}
\usepackage{amsmath}
\usepackage{amssymb}
\usepackage{dsfont}
\usepackage{physics}
\usepackage{hyperref}
\usepackage{tikz}
\usetikzlibrary{calc,positioning,shapes.geometric,arrows.meta}
\tikzset{
  arr/.style={
    -{Stealth[length=1.6mm,width=1.6mm]},
    line width=0.8pt,
    line cap=round,
    line join=round
  }
}

\DeclareMathOperator{\sign}{sign}
\newcommand{\vecK}{\mathbf{K}}
\newcommand{\vecq}{\mathbf{q}}
\newcommand{\veck}{\mathbf{k}}
\newcommand{\vecQ}{\mathbf{Q}}
\newcommand{\vecM}{\mathbf{M}}
\newcommand{\vecr}{\mathbf{r}}
\newcommand{\vecR}{\mathbf{R}}
\newcommand{\vech}{\mathbf{h}}
\newcommand{\veca}{\mathbf{a}}
\newcommand{\vecd}{\bm{\delta}}
\newcommand{\vecs}{\bm{\sigma}}

\newcommand{\rn}[1]{{\color{purple}#1}}

\begin{document}

\title{Impurity as a probe of Berry curvature and wavefunction winding in gapped two-band models}

\author{Reda Nabil}
\author{Pascal Simon}
\author{Andrej Mesaros}
\affiliation{Universit\'e Paris-Saclay, CNRS, Laboratoire de Physique des Solides, 91405, Orsay, France}

\date{\today}

\begin{abstract}
We explore the possibility of using quasiparticle interference near impurities to extract non-local properties of bands. Focusing on the minimal two-band description of a system with multiple valleys, we show that the local density of states (LDOS) induced by non-magnetic impurities can effectively probe both the Berry curvature and the wavefunction winding, locally in each valley. We analyze gapped models based on Dirac, semi-Dirac, quadratic touching, and higher-order touching energy dispersions, in presence of a point-like potential scatterer. We find that for strong enough impurity potentials the LDOS along a contour around the impurity unambiguously shows the wavefunction winding number, in contrast to the case of gapless Dirac-like dispersions where the winding can be precluded by LDOS oscillations in the radial direction. We also establish a general connection between, on the one hand, the Berry curvature as a function of momentum locally in a valley, and on the other hand, the wavefunction winding number and the energy dispersion parameters, making hence the local Berry curvature directly observable from spectroscopic LDOS information.
\end{abstract}

\maketitle

\section{Introduction}
The theory of quasiparticle interference (QPI)\cite{Hoffman2002,Wang2003,Balatsky2006} uses the LDOS induced by an impurity to characterize the bulk substrate. It has been established that QPI can reveal information about the substrate's wavefunction that is not apparent in its energy dispersion, for example, LDOS measured by scanning tunneling spectroscopy (STS) was used to probe the pairing function of superconductors\cite{Pereg-Barnea2008,Nunner2006,Bhattacharyya2023,Chi2017,Chen2017,SeamusDwave,SeamusFwave,Hirschfeld2021}. In the case of a two-valley Dirac semimetal such as graphene, the LDOS response to impurity was studied early on\cite{Wehling2007,Pereg-BarneaMacDonald2008,Bena2008-hs,Wehling2009}, while only recently it was discovered\cite{Dutreix2019} that the inter-valley interference reveals a non-local quantity: the pseudo-spin winding number $W_\pm=\pm1$ of the wavefunction around the Dirac point $\pm\vecK$ appears at the impurity as a vortex of strength $W_+-W_-=2$ (in units of $2\pi)$ in the phase of the real-space LDOS oscillations. This phenomenon appears generally in semimetals and nodal superconductors protected by chiral symmetry, but is ambiguous and requires tuning of the impurity and the STS tip\cite{LenaCones}. Namely, in the LDOS the interference terms exhibiting the dislocation compete with the ones that do not, since the amplitudes of these terms show Friedel oscillations with period $\hbar v_D/E$ in the radial direction, $E$ being the energy and $v_D$ the Dirac velocity. Hence, instead of a single dislocation in the LDOS at the impurity, there generically appear dislocation-antidislocation pairs at various distances to it. An exception occurs for a vacancy defect, which is pinned at $E=0$ by chiral symmetry, so there are no oscillations and the single dislocation strikingly shows the winding number\cite{Akkermans2023,Akkermans2025}. In contrast to this understanding of nodal systems, it remains an open question if some non-local and topological properties of the wavefunction of a gapped system can be extracted using simple impurities.

Experimental motivation comes from the fact that the lowest energy excitations in transition metal dichalcogenide (TMD)\cite{Xu2014-go} monolayers have been observed to have a Dirac nature 
\cite{Aivazian2015,Wang2017}, in accordance with the theoretical description as Dirac electrons with multiple gapping parameters\cite{Xiao2012,Bieniek2018}. TMDs are well-known to have many types of defects, some of which have already been studied by STS\cite{Sheina2023}. Another interesting system are the Dirac surface modes of a topological insulator that may be both scattered and gapped by magnetic impurities\cite{Roushan2009-ge,black2015}, and it would be of interest to further characterize the Dirac nature of these modes based on STS data.

Conceptually, two questions arise when one considers gapped valleys. First, the impurity may bind an in-bulkgap state, which would hence be exponentially localized in real space, and possibly lacking Friedel oscillations, bringing into question the observability of dislocation(s). Second, the Dirac wavefunction has two quantized properties, the Berry phase $\pm\frac{1}{2}$ (in units of $2\pi$) due to a singular Berry curvature\cite{Castro_Neto2009-wu} at $\pm\bm{K}$, and the above-mentioned pseudo-spin winding number $W_\pm$. These are tied to each other, historically sometime confounded \cite{Park2011,Novoselov2006}. Once the gap opens, however, the Berry curvature spreads throughout the valleys, and the Berry phase is non-quantized, while the winding still is. The question is then if the LDOS around the impurity can reveal both of these independent non-local quantities.

In this work, we consider several two-band, multi-valley models with a gap, having various winding numbers and low-energy dispersions, all derived from the honeycomb lattice. We find that a strong enough on-site potential scatterer binds, inside the bulkgap, a bound state that exhibits a single wavefront dislocation at the impurity, making the winding number robustly observable. We then consider a general two-band model in which each valley has an isotropic low-energy dispersion with leading order $|\vecq|^n$ momentum dependence, and we show that the extracted winding number $W$ together with the spectroscopically accessible parameters of the dispersion reveal the Berry curvature $\Omega(|\vecq|)$ in the valley. To leading order, we find $\Omega(|\vecq|)\propto W |\vecq|^{2(n-1)}$. We also comment on how the functional form of $\Omega(\vecq)$ depends on a winding-number density in the case of anisotropic dispersions. Overall, somewhat surprisingly, our results show that STS data on simple (strong) potential scatterers in small-gap semiconductors should provide direct access to both wavefunction winding and Berry curvature.

This paper is organized as follows: In Section~\ref{sec:TB} we introduce a tight-binding honeycomb model with tuned farther neighbor hoppings producing several continuum models, as well as a more general two-band isotropic model. In Section~\ref{sec:wb_fV} we solve the T-matrix problem for these continuum models finding the bound states and the form of the LDOS. In Section~\ref{ssec:Winding regions} we study the observability of the wavefront dislocation in the LDOS. In Section~\ref{sec:Berry curvature} we use a small momentum expansion of a two-band model for a valley to relate the Berry curvature function to the winding number and the parameters of the energy dispersion, applying this to each of the considered isotropic continuum models. In Section~\ref{sec:Hybrid} we consider a semi-Dirac anisotropic valley model, and relate the dipole form of the berry curvature to the winding behavior of the pseudospin of the wavefunction. We close with a discussion and conclusions in Section~\ref{sec:conclusions}.

\section{The tight-binding models}
\label{sec:TB}
\begin{figure}[h!]
\begin{tikzpicture}[
    scale=1.2,
    A/.style={circle, draw=gray, fill=white, inner sep=2.4pt},
    B/.style={circle, fill=black, inner sep=2.4pt},
    imp/.style={regular polygon, regular polygon sides=3, draw=green!60!black, fill=green!30, inner sep=1.2pt, rotate=0},
    vec/.style={->, thick},
]
\def\a{1.3}
\coordinate (a1) at ({sqrt(3)/2*\a},{-0.5*\a});
\coordinate (a2) at ({sqrt(3)/2*\a},{ 0.5*\a});
\coordinate (d1) at ({-sqrt(3)/3*\a},0);
\coordinate (d2) at ({ sqrt(3)/6*\a},{-0.5*\a});
\coordinate (d3) at ({ sqrt(3)/6*\a},{0.5*\a});
\foreach \m/\n in {0/0, 1/0, 0/1, 1/1, -1/0, 0/-1} {
    \coordinate (A\m\n) at ($\m*(a1)+\n*(a2)$);
}
\foreach \m/\n in {0/0, 1/0, 0/1, 1/1} {
    \coordinate (B\m\n) at ($(A\m\n)+(d1)$);
}
\foreach \m/\n in {0/0, 1/0, 0/1, 1/1, -1/0, 0/-1} {
    \draw[gray!65] (A\m\n) -- ($(A\m\n)+(d1)$);
    \draw[gray!65] (A\m\n) -- ($(A\m\n)+(d2)$);
    \draw[gray!65] (A\m\n) -- ($(A\m\n)+(d3)$);
}
\draw[gray!65] (B10) -- ($(B10)-(d3)$);
\draw[gray!65] (B01) -- ($(B01)-(d2)$);
\foreach \m/\n in {0/0, 1/0, 0/1, 1/1, -1/0, 0/-1} {
    \node[A] at (A\m\n) {};   
}
\foreach \m/\n in {0/0, 1/0, 0/1, 1/1} {
    \node[B] at (B\m\n) {};
}
\draw[vec, blue] (A00) -- ($(A00)+(a1)$)
    node[pos=0.8, above] {$\vec a_1$};
\draw[vec, blue] (A00) -- ($(A00)+(a2)$)
    node[pos=0.8, below] {$\vec a_2$};
\draw[vec, red] (A00) -- ($(A00)+(d1)$)
    node[pos=0.8, below] {$\vec\delta_1$};
\draw[vec, red] (A00) -- ($(A00)+(d2)$)
    node[pos=0.8, below left] {$\vec\delta_2$};
\draw[vec, red] (A00) -- ($(A00)+(d3)$)
    node[pos=0.8, above left] {$\vec\delta_3$};
\node[imp] (V) at (A00) {};
\node[green!60!black, above=1pt of V] {$V$};
\node[A, label=right:{A}] at (3.4,1.2) {};
\node[B, label=right:{B}] at (3.4,0.8) {};
\node[imp, label=right:{impurity}] at (3.4,0.4) {};
\node[right] at (3.1,-0.6)
{$\vec a_1=\left(\frac{\sqrt3}{2},-\frac{1}{2}\right)a$};
\node[right] at (3.1,-1.1)
{$\vec a_2=\left(\frac{\sqrt3}{2},+\frac12\right)a$};
\end{tikzpicture}
\caption{Notation for the honeycomb lattice model. Each unit-cell is formed of an atom A and an atom B shifted by $\vecd_1$ from A. The Bravais lattice is spanned by the vectors $\mathbf{a}_1$ and $\mathbf{a}_2$ taken to be of unit length $a=1$. The impurity term $\hat{V}$ is localized on a site A, modifying its potential energy by $V$.}
\label{fig:lattice_graphene}
\end{figure}
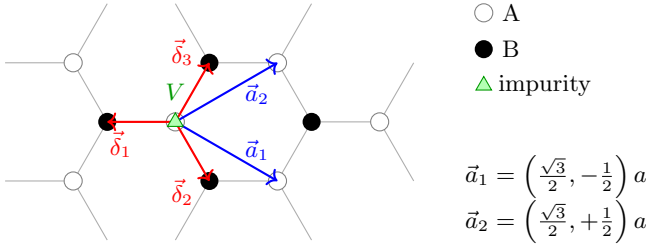

We start with a honeycomb lattice structure defined in Fig.~\ref{fig:lattice_graphene}, on which we will introduce hoppings to build various models in the following subsections.

\subsection{Spinless Semenoff mass model}
\label{ssec:TB}

The first model is the standard hBN \cite{Semenoff1984-od}, with nearest-neighbor (NN) hoppings and on-site energies $(M,-M)$:
\begin{equation}
\begin{aligned}
H_0^{\textrm{hBN}}
={}&
t_1 \sum_{\vecR}\sum_{i=1}^{3}
\left(
c_{\vecR+\vecd_i,B}^{\dagger}
c_{\vecR,A}^{\phantom{\dagger}}
+\mathrm{h.c.}
\right) \\
&+
M\sum_{\vecR}
\left(
c_{\vecR,A}^{\dagger}c_{\vecR,A}^{\phantom{\dagger}}
-
c_{\vecR,B}^{\dagger}c_{\vecR,B}^{\phantom{\dagger}}
\right),
\end{aligned}
\label{eq:H_hBN}
\end{equation}
with the electron annihilation operator $c_{\vecR,\alpha}$ on sublattice $\alpha=A,B$. This gives the two-by-two Bloch Hamiltonian matrix:
\begin{equation}
H_0^{\text{hBN}}(\veck) = 
\begin{pmatrix}
M & t_1 f_1(\veck)^* \\
t_1 f_1(\veck) & -M
\end{pmatrix},
\label{eq:H0_hBN}
\end{equation}
where we introduce the notation $f_1(\veck)$ for the NN structure function of graphene $f_1(\veck)\equiv \sum_{i=1,2,3} e^{+i \veck \cdot \vecd_i}$, see Appendix~\ref{sec:structure_factor_expansions}. Linearization of Eq.~\eqref{eq:f1_K_expansion} at each valley $\vecK_\xi = 2 \pi(\frac{1}{\sqrt{3}},- \xi \frac{1}{3})$, with $\xi=\pm1$, gives
\begin{equation}
H_{0}^{\textrm{hBN}}(\vecK_\xi + \vecq) = 
\begin{pmatrix}
M & v^* |\vecq| e^{+i \xi \theta_q} \\
 v |\vecq| e^{-i \xi \theta_q}  & -M 
\end{pmatrix},
\label{eq:linearized_hBN}
\end{equation}
where $v = e^{i\frac{5\pi}{6}} \frac{\sqrt{3}}{2}t_1$ is the Fermi velocity $v_F$ up to a complex prefactor related to the gauge we have chosen, and $\theta_q$ is the polar angle of the small momentum $\vecq$ in the valley. We emphasize that the off-diagonal components of the Hamiltonian present a $\xi 2\pi$-pseudospin winding along a momentum loop which is located within the valley $\xi$ and encloses the $\vecK_\xi$ Dirac point.

\subsection{Spin-orbit coupled model with Kane-Mele and Valley-Zeeman terms}
\label{ssec:SOC}
In order to study the massive Dirac fermions relevant for TMDs, where Berry curvature and spin-valley effects play a central role \cite{Xiao2012,Bieniek2018}, we introduce spin-orbit coupling (SOC) parameters: $t_{KM}$ and $t_{VZ}$, corresponding to Kane-Mele \cite{Kane2005-cf} and valley-Zeeman \cite{McCann12,Wang16} SOC,  respectively. Their contributions to the Hamiltonian are (see Appendix~\ref{app:SOC_conventions} for detailed conventions):
\begin{equation}
H^{\textrm{SOC}} = i \sum_{\vecR} \sum_{s=\pm} \sum_{i=1}^3 \sum_{\alpha = A,B} s \lambda_i^\alpha c_{\vecR+\vecd_i',\alpha,s}^\dagger c_{\vecR,\alpha,s}^{\phantom{\dagger}} + \text{h.c.}
\end{equation}
where $s$ represents the spin, and
\begin{align}
&\lambda^A = \left( t_{VZ} + t_{KM} \right)
\begin{pmatrix}
\phantom{-}1 \\
-1 \\
\phantom{-}1
\end{pmatrix}\\
&\lambda^B = \left( t_{VZ} - t_{KM} \right)
\begin{pmatrix}
\phantom{-}1 \\
-1 \\
\phantom{-}1
\end{pmatrix},
\end{align}
giving, together with the hBN terms in Eq.~\eqref{eq:H0_hBN}, the total Bloch Hamiltonian block for spin $s$:
\begin{equation}
H_{0,s}^{\textrm{SOC}}(\veck) = s t_{VZ} F_\veck \sigma_0 + 
\begin{pmatrix}
M + s t_{KM} F_\veck & t_1 f_1^* \\
t_1 f_1  & -M - s t_{KM} F_\veck
\end{pmatrix},
\end{equation}
where the form function $F_\veck$ is defined in Appendix~\ref{app:socF}. The linearized valley Hamiltonian is then
\begin{equation}
H_{0,s}^{\textrm{SOC}}(\vecK_\xi + \vecq) = s \xi \tilde t_{VZ}  \sigma_0 + 
\begin{pmatrix}
M + s \xi \tilde t_{KM} & v^* |\vecq| e^{i \xi \theta_q} \\
v |\vecq| e^{-i \xi \theta_q}  & -M - s \xi \tilde t_{KM} 
\end{pmatrix},
\label{eq:linearized_KM}
\end{equation}
with $\tilde t_{VZ}\equiv3 \sqrt{3} t_{VZ}$ and $\tilde t_{KM}\equiv3 \sqrt{3} t_{KM}$.

The three coupling terms compete \cite{Fabian17}, and the resulting gap for each spin $s$ at each $\vecK_\xi$ valley covers the range of energies:
\begin{equation}
E\in\left[ s \xi \tilde t_{VZ} - |M + s \xi \tilde t_{KM}|, \, s \xi \tilde t_{VZ} + |M + s \xi \tilde t_{KM}| \right].
\end{equation}
In the continuum limit, we will obtain impurity-bound states by summing the propagator over all valleys, so we require that the bound-state energy is within the spectral gap of all valleys simultaneously.

In the remainder of this work, we focus on two representative parameter sets for the spinful models. First, we consider the pure Kane--Mele (KM) model, where $M$ and $t_{VZ}$ are both set to $0$.

Second, we consider a parameterization representative of the transition metal dichalcogenide (TMD) $\text{MoS}_2$, taken from Ref.~\cite{Xiao2012} \rn(units are fixed by setting $a=\hbar=1$),
\begin{equation}
M=0.83,\quad
t_1=5.06,\quad
t_{\mathrm{VZ}}=-\tilde t_{\mathrm{KM}}=0.04.
\label{param_MoS2}
\end{equation}

\subsection{Spinless quadratic and semi-Dirac models}
\label{ssec:Quadratic&hybrid}
To access higher winding numbers and discuss anisotropic valleys, we use NNN ($t_2$) and NNNN ($t_3$) hoppings, as introduced previously \cite{De_Gail2012-pk,Montambaux2009-yn,Montambaux2009-nv}. The hBN Hamiltonian with these added hoppings takes the following Bloch form:
\begin{equation}
H_0^{\text{NNNN}}(\veck) = 
\begin{pmatrix}
M + t_2 f_2(\veck) & t_1 f_1(\veck)^* + t_3 f_3(\veck)^* \\
t_1 f_1(\veck) + t_3 f_3(\veck) & -M - t_2 f_2(\veck)
\end{pmatrix},
\label{eq:H0_NNNN}
\end{equation}
where the expressions of form factors $f_i(\veck)$ are in Appendix~\ref{sec:structure_factor_expansions}.

\subsubsection{Quadratic model}
\label{sssec:Quadratic}
Finely tuning the NNN and NNNN hoppings in Eq.~\eqref{eq:H0_NNNN} gives quadratic band touchings at $\vecK_\xi$ that get gapped and carry a $\xi 4\pi$-pseudospin winding number -- doubled compared to hBN. Concretely, we set $t_2=0$ and $t = t_1 = 2t_3$ and linearize the Bloch Hamiltonian:
\begin{equation}
H_0^{\textrm{quad}}(\vecK_\xi+\vecq) = 
\begin{pmatrix}
M   & \beta^* |\vecq|^2 e^{-i 2 \xi \theta_q} \\
\beta |\vecq|^2 e^{i 2 \xi \theta_q}  & -M
\end{pmatrix},
\label{eq:linearized_quad}
\end{equation}
with $\beta\equiv e^{i\frac{\pi}{3}} \frac{3}{8} t$.

\subsection{General valley Hamiltonian}
\label{ssec:General}
All the above two-by-two Bloch Hamiltonian matrices can be written in the usual way in terms of the Pauli matrices $\vecs\equiv(\sigma_x,\sigma_y,\sigma_z)$ and a Bloch vector $\vech(\veck)\equiv(h_x(\veck),h_y(\veck),h_z(\veck))$:
\begin{equation}
H_0(\veck) = h_0(\veck) \mathbb{1}+\vech(\veck)\cdot\vecs.
\label{eq:vech}
\end{equation}

Considering now the low-energy valley expansions, all the isotropic models can be captured by the following continuum Hamiltonian:
\begin{equation}
H_0(\vecQ + \vecq) = h_{0 \vecQ} \mathbb{1} +
\begin{pmatrix}
\Delta_Q & \lambda^* |q|^n e^{-iW_Q\theta_q} \\
\lambda |q|^n e^{+iW_Q\theta_q} & -\Delta_Q
\end{pmatrix},
\label{linearised_general_H}
\end{equation}
where $\Delta_Q$ is the gap at the valley $\vecQ$, the $\lambda$ is a generally complex prefactor, the off-diagonal term carries the winding number $W_Q$ of the valley, while the low-energy dispersion scales as $|\vecq|^n$. Precisely, this continuum description encompasses the cases: $n=1$ with $W=\pm1$ for hBN (Eq.~\eqref{eq:linearized_hBN}) and SOC (Eq.~\eqref{eq:linearized_KM}) models, and $n=2$ with $W=\pm2$ for the quadratic model (Eq.~\eqref{eq:linearized_quad}). However, we will consider the valley model in Eq.~\eqref{linearised_general_H} generally, with independent parameters $(n,W_Q,\lambda,\Delta_{Q})$.

\section{Bound states}
\label{sec:wb_fV}
Starting from any bulk model considered in the previous section, the impurity is considered as a local scattering potential $V$ on sublattice $A$ of the lattice unit-cell $\vecR=\mathbf{0}$, thus locally breaking the chiral sub-lattice symmetry.
The impurity Hamiltonian term reads:
\begin{equation}
    \hat{V} = V \text{ } c_{\mathbf{0},A}^\dagger c_{\mathbf{0},A}^{\phantom{\dagger}}=\frac{V}{N}
\sum_{\mathbf{k},\mathbf{k}'}
c_{\mathbf{k},A}^\dagger
c_{\mathbf{k}',A}^{\phantom{\dagger}},
\end{equation}
for a crystal containing $N$ unit cells.

\subsection{LDOS due to the impurity}
\label{ssec:LDOS}
For any choice of bulk model in presence of the impurity, we focus on the impurity-induced bound state(s) within the bulkgap of the valley. We hence use the $T$-matrix approach \cite{Shiba1968,Pereg-BarneaMacDonald2008,Balatsky2006} (see also Appendix~\ref{sec:Greens} for details of our calculation), which gives the impurity-induced correction to the LDOS with respect to the original bulk value of LDOS $\rho_0(\mathbf r,\omega)$:
\begin{equation}
    \delta \rho(\mathbf r,\alpha;\omega)
    =
    -\frac{1}{\pi}
    \Im
    \left[
    G(\mathbf r,\mathbf r;\omega)_{\alpha\alpha}
    -
    G_0(\mathbf r,\mathbf r;\omega)_{\alpha\alpha}
    \right],
\label{delta_rho1}
\end{equation}
where $G$ denotes the Green's function in the presence of the impurity, $G_0$ that of the bulk crystal, and $\alpha$ still labels the sublattice degree of freedom.

For energies lying inside the bulkgap, the impurity correction becomes
\begin{align}
    \delta \rho(\mathbf r,\alpha;\omega)
    \propto
    \delta(\omega-\omega_b)
    \Re
    \left[
    G_0(\mathbf r,\mathbf 0;\omega)_{\alpha A}
    G_0(\mathbf 0,\mathbf r;\omega)_{A\alpha}
    \right],
\label{eq:LDOS}
\end{align}
where $\omega_b$ denotes the bound-state energy, whose value we will determine later.

Using the explicit expression for the two-by-two Green's function of the Bloch Hamiltonian, as in Eq.~\eqref{eq:vech}, we can get the impurity-induced LDOS of the lattice model on each sublattice:
\begin{align}
&\delta \rho_A(\vecr, \omega_b) \propto \left| \sum_{\veck} e^{-i \veck \cdot \vecr} \frac{\tilde \omega_b + h_z}{|\vech |^2 - \tilde \omega_b^2} \right|^2\\
&\delta \rho_B(\vecr, \omega_b) \propto \left| \sum_{\veck} e^{-i \veck \cdot \vecr} \frac{h_x + i h_y}{|\vech|^2 - \tilde \omega_b^2} \right|^2.
\end{align}
where $\tilde \omega_b = \omega_b - h_0$.
Notice that the position of the local impurity on the A sublattice causes a different response in the two sublattices.
We will use the tight-binding results as a benchmark against the low-energy valley Hamiltonian results. We proceed by finding the bulk Green's function $G_0(\mathbf r,\mathbf 0;\omega)$ for the general valley Hamiltonian in Eq.~\eqref{linearised_general_H}, which can represent any of the honeycomb-based valley models we introduced above.

\subsection{Green's function and LDOS for the general valley Hamiltonian}
\label{ssec:continuum_green}
Surprisingly, the bulk real-space Green's function of the general valley Hamiltonian in Eq.~\eqref{linearised_general_H} has an analytical closed form expression, which we obtain by Fourier transforming to real space the continuum variable $\vecq$, yielding:
\begin{widetext}
\begin{equation}
G_0(\vecr,0;\omega) \approx
\frac{2\pi}{A_{\mathrm{BZ}}}
\sum_{\vecQ}
\frac{e^{-i\vecQ\cdot\vecr}}
{n|\lambda|^{2/n}\Omega^{2(1-1/n)}}
\sum_{j=0}^{n-1}
\begin{pmatrix}
(\tilde\omega+\Delta_{Q})
\, \zeta_j^2
K_0(-i\zeta_j\tilde r)
&
\frac{(-1)^{W}\lambda^*}{|\lambda|}  
e^{-iW\theta_{r}}
\Omega \, \zeta_j^{*(n +2)}
K_W(+i\zeta_j^*\tilde r)
\\
\frac{(-1)^{W}\lambda}{|\lambda|} 
e^{iW\theta_{\vecr}}
\Omega \, \zeta_j^{(n +2)}
K_W(-i\zeta_j\tilde r)
&
(\tilde\omega-\Delta_{Q})
\, \zeta_j^2
K_0(-i\zeta_j\tilde r)
\end{pmatrix},
\label{eq:Green}
\end{equation}
\end{widetext}
where the matrix is written in the $(A,B)$ sublattice basis, and we simplified notation by dropping the index of $W_Q$. We introduced
\begin{equation}
\zeta_j
=
\exp\left(
i\frac{2j+1}{4n}2\pi
\right),
\end{equation}
$\Omega=\sqrt{\Delta_{Q}^2-\tilde\omega^2}$,
$\tilde r=\left(\Omega/|\lambda|\right)^{1/n}|\vecr|$, $\tilde \omega \equiv \omega - h_0(\vecQ)$, $A_{\mathrm{BZ}}=\frac{8\sqrt3}{3}\pi^2$. Note that the parameters $n$, $W$, $\Delta$, $\lambda$, and the variables $\tilde{\omega}$, $\Omega$, $\tilde{r}$ all depend on the choice of the valley $Q$, but we don't label this in order to lighten notation. Finally, the $\theta_r$ is the real-space polar angle with impurity at the origin, and $K_m$ is the modified Bessel function of the second kind of order $m$. Since the numbers $\zeta_j$ are in the upper complex plane, the arguments of the Bessel functions have a positive real part, and hence decay exponentially.

Equation~\eqref{eq:Green} provides a unified expression for the real-space propagator of all the isotropic low-energy Hamiltonians considered in this work, but can also be applied for any other values of parameters $(n,W,\lambda,\Delta)$. This is one of the most important analytical results of this paper.

The Green's function, Eq.~\eqref{eq:Green}, is a sum over valleys $\vecQ$, hence each term in the LDOS, Eq.~\eqref{eq:LDOS}, represents a scattering from a valley, say $\vecQ_1$ to another valley, say $\vecQ_2$. The intra-valley terms have $\Delta\vecQ\equiv\vecQ_1-\vecQ_2=0$, while inter-valley contributions have $\Delta\vecQ=\vecQ_1-\vecQ_2\neq0$. Focusing on a specific inter-valley contribution $\Delta \vecQ \neq0$, we have:
\begin{align}\delta \rho (\Delta \vecQ; \vecr,\omega) =& \delta \rho_A(|\vecr|, \omega) \cos(\Delta \vecQ\cdot \vecr)\notag \\
+& \delta \rho_B(|\vecr|,\omega) \cos(\Delta \vecQ\cdot \vecr + \Delta \varphi + \Delta W\theta_r),
\label{eq:dAvsdB}
\end{align}
where $\Delta W\equiv W_{Q_1}-W_{Q_2}$ is the relative winding number of the two valleys, while $\Delta\varphi = \arg(\lambda_{Q_1}) - \arg(\lambda_{Q_2})$ is a constant phase shift. We have added together the LDOS on both sublattices, having in mind that in real STS data it is hard to achieve high real-space resolution around the impurity.

Here we highlight two new features of the impurity-induced LDOS compared to previous work on gapless Dirac systems which had $n=1$, $\Delta W=\pm1$. First, we find that the pseudospin winding numbers $W_Q$ of the Bloch Hamiltonian are transferred directly into the inter-valley terms of the LDOS, Eq.~\ref{eq:dAvsdB}, even for higher values of the $|W_Q|>1$. Second, 
we use the particular example of the hBN model, which has the amplitudes $\delta \rho_A(|\vecr|,\omega) \propto (\omega + M)^2 K_0^2(|\vecr|\Omega/v_F)\delta(\omega-\omega_b)$ and $\delta \rho_B(|\vecr|,\omega) \propto (M^2 - \omega^2) K_1^2(|\vecr| \Omega / v_F)\delta(\omega-\omega_b)$, to emphasize that the power-law decaying and oscillating Hankel functions of the gapless Dirac case have been replaced by exponentially decaying and monotonous modified Bessel functions, reflecting the localization of the in-bulkgap bound state and absence of Friedel oscillations. We will see that these properties of the amplitudes lead to a more robust observability of $W_Q$ in the LDOS, compared to the case of gapless Dirac systems.

\subsection{Bound-state condition}
\label{ssec:wb_fV}

So far, we have assumed that there is a bound state at energy $\omega_b$. The bound state actually exists if there is a pole of the full Green's function inside the bulkgap (i.e., the gap of the $H_0$). Within the $T$-matrix formalism, this condition is equivalent to requiring $\Re\det\!\left[1-\hat{V}\hat{G}_0(\omega_b)\right]\equiv0$, which determines the bound-state energy $\omega_b$. For our real impurity potential on an $A$ site the condition simplifies to
\begin{equation}
1-V\Re\!\left[G_0(\mathbf 0,\mathbf 0;\omega_b)_{AA}\right]\equiv0.
\label{eq:bs_energy_condition}
\end{equation}

For a two-by-two Green's function, one can rewrite this equation exactly as
\begin{equation}
\frac{1}{V}=-\frac{1}{N}\sum_{\veck}
\frac{\tilde\omega_b+h_z}{|\vech|^2-\tilde\omega_b^2},
\label{eq:linear_time_bs_energy}
\end{equation}
which provides an efficient numerical method for determining the exact bound-state energy within the tight-binding lattice model, with a computational cost scaling linearly with the number of unit cells.

In the low-energy theory, Eq.~\eqref{eq:Green} yields closed analytical expressions for the bound-state energy as a function of the impurity strength. These expressions fall into two universality classes depending on the low-energy dispersion: Dirac systems with linear dispersion ($n=1$) and systems with higher-order band touching ($n>1$). We illustrate these results using four representative materials: hBN, the Kane--Mele model, monolayer $\text{MoS}_2$, and the quadratic model.

\subsubsection{Bulk linear Dirac systems ($n=1$)}
\label{sssec:wb_linear}

The low-energy bulk Hamiltonians of hBN, the Kane--Mele model, and monolayer MoS$_2$ all exhibit a linear Dirac dispersion around each valley. Applying Eq.~\eqref{eq:Green} to these systems yields the bound-state condition
\begin{equation}
\frac{1}{V}
=
-\frac{2\pi}{A_{\mathrm{BZ}}}
\sum_{\vecQ}
\cos(\vecQ\!\cdot\!\veca_0)
\frac{\tilde\omega_b+\Delta_{\vecQ}}
{|\lambda|^2}
K_0\!\left(
\frac{\Omega_b|\veca_0|}{|\lambda|}
\right),
\label{eq:bs_linear_general}
\end{equation}
where
\(
\Omega_b=\sqrt{\Delta_{\vecQ}^2-\tilde\omega_b^{\,2}}
\),
and the sum runs over all inequivalent valleys.

Unlike the case $n>1$, the continuum Green's function diverges logarithmically at the origin. We therefore introduce a short-distance cutoff through the vector $\veca_0$, whose magnitude is of the order of the lattice spacing and below which the continuum approximation breaks down. Its length and orientation are obtained by fitting the analytical expression to the exact tight-binding result from Eq.~\eqref{eq:linear_time_bs_energy}.

Fig.~\ref{fig:wb_fV_all}a,b,c compares the resulting analytical predictions with tight-binding calculations for hBN, the Kane--Mele model and monolayer MoS$_2$.

\subsubsection{Bulk higher-order band-touching systems ($n>1$)}
\label{sssec:wb_higher}

For low-energy bulk Hamiltonians whose dispersion scales as $q^n$ with $n>1$, the momentum integral converges at short distances and no ultraviolet regularization is required. The bound-state condition becomes
\begin{equation}
\frac{1}{V}
=-
\frac{\pi}{A_{\mathrm{BZ}}}
\sum_{\vecQ}
\frac{\pi/n}
{\sin(\pi/n) \, |\lambda|^{2/n}\,
}
\,
\frac{\tilde\omega_b+\Delta_{\vecQ}}{\Omega_b^{\,2(1-1/n)}},
\label{eq:bs_higher_general}
\end{equation}
with
\(
\Omega_b=\sqrt{\Delta_{\vecQ}^2-\tilde\omega_b^{\,2}}
\).

\begin{figure}[h!]
\includegraphics[width=0.49\textwidth]{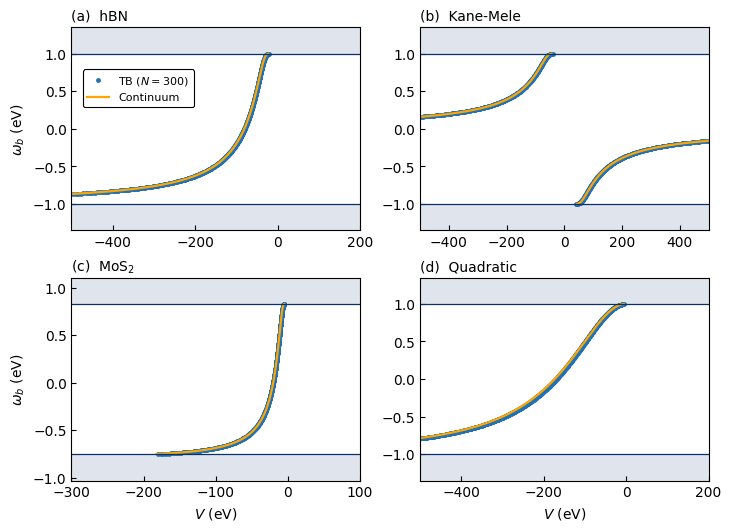}
\caption{Bound-state energy $\omega_b$ as a function of the impurity strength $V$ for four gapped systems: hBN [$M=1$, $t_1=10$], Kane--Mele (KM) [$\tilde t_{KM}=1$, $t_1=10$], MoS$_2$ [see parameters from eq.~\eqref{param_MoS2}], and the quadratic model [$M=1$, $t=100$]. Blue dots show tight-binding results for a $300\times300$-unit-cell system, while solid orange lines correspond to the analytical continuum theory. For the three linear models (hBN, KM, and MoS$_2$), the continuum results are fitted using the cutoff parameter $\vec a_0=(0,a/3)$, whereas the quadratic model requires no fitting parameter. The bound-state energy is restricted to the bulk gap. For hBN and the quadratic model, a bound state exists for values of $V$ of the opposite sign to $\Delta_\vecQ$, i.e., over one half of the real $V$ axis. For MoS$_2$, bound states occur only within a finite interval $[V_{\min},V_{\max}]$. In contrast, for the KM model, any nonzero impurity strength $V\neq0$ gives rise to a bound state.}
\label{fig:wb_fV_all}
\end{figure}

Equation~\eqref{eq:bs_higher_general} applies to all higher-order band-touching models and is illustrated in this work for the quadratic model ($n=2$). Fig.~\ref{fig:wb_fV_all}d compares this prediction with the exact tight-binding calculation for the quadratic model.

Interestingly, we can consider the limit of flat bands in the valleys by taking the dispersion exponent to infinity, $n\rightarrow\infty$. We then find
\begin{equation}
\frac{1}{V}
\longrightarrow -
\frac{\pi}{A_{\mathrm{BZ}}}
\sum_{\vecQ}
\frac{1}
{\Delta_{\vecQ}-\tilde\omega_{b,\infty}}.
\end{equation}
In the case where $\Delta_{\vecQ}+h_0(\vecQ)$ is independent of $\vecQ$, we recover a linear relation between the impurity strength and the bound-state energy, reminiscent of the behavior found in general flat-band systems:
\begin{equation}
\omega_{b,\infty}
=(\Delta+h_0)
+\frac{N_{\vecQ}\pi}{A_{\mathrm{BZ}}} \,V,
\end{equation}
where $N_{\vecQ}$ denotes the number of valleys included in the low-energy description.

\section{Observability of the winding numbers}
\label{ssec:Winding regions}
\begin{figure}[h!]
\includegraphics[width=0.5\textwidth]{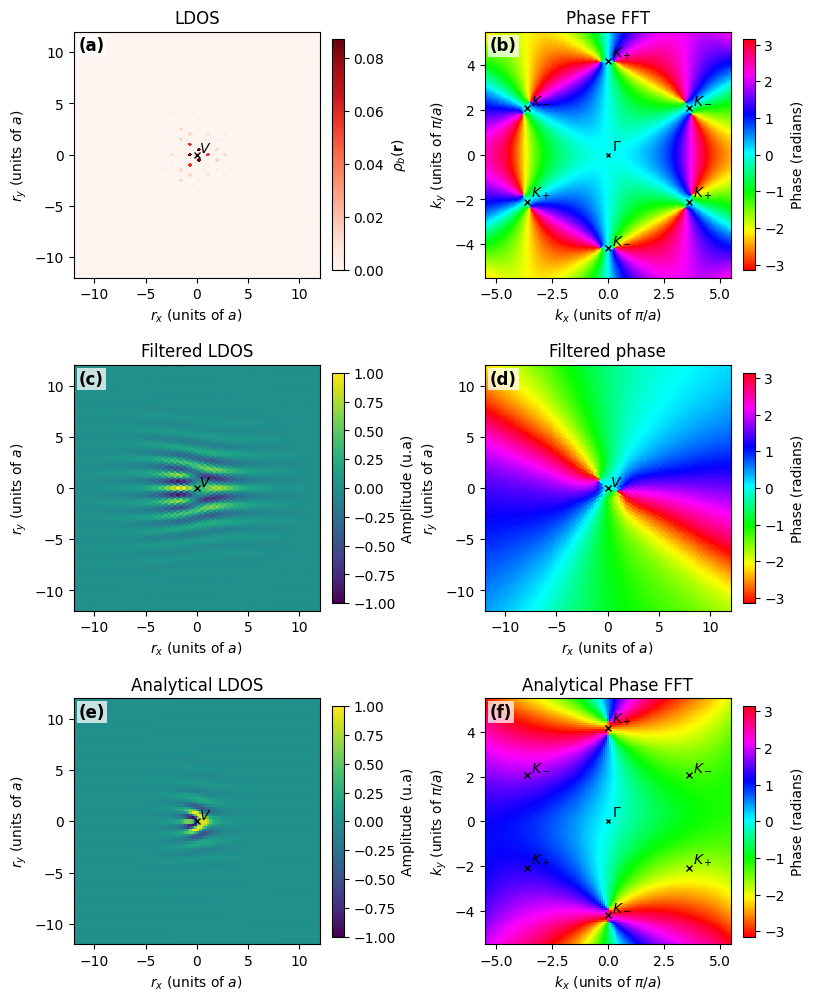}
\caption{Detecting the winding number in LDOS for hBN model. (a) LDOS around the impurity in the tight-binding model with parameters $t_1 = 10M =10$, $V=-80M$ leading to a bound state in the middle of the gap $\omega_b = 0.0M$. (b) Phase of the Fourier Transform of the LDOS in (a), showing the $\pm4\pi$ vortices at the $\vecK$-points. (c) Amplitude of the  LDOS in (a) after applying a Gaussian filter to extract modulations near the $\vecK$-points, showing a wavefront dislocation of strength $2$ (see text). (d) Phase of the filtered LDOS after applying a Gaussian filter on the $\vecK_+$ valley, shows a $-4\pi$-winding. (e) LDOS from analytical low-energy theory, showing the dislocation. (f) Phase of the Fourier transform of (e), shows $\pm4\pi$ vortices around the $\vecK$-points.}
\label{fig:filtered_LDOS}
\end{figure}

We now illustrate the main results on the representative case of the hBN model, comparing the LDOS of the tight-binding and the low-energy theory. Fig.~\ref{fig:filtered_LDOS}a shows the raw tight-binding output for the LDOS. The relative winding number $\Delta W\equiv W_{K_+}-W_{K_-}=2$ becomes visible in the Fourier transform of the LDOS (Fig.~\ref{fig:filtered_LDOS}b), as a vortex of $2\pi\Delta W=4\pi$ in the phase located at $\Delta\vecK=2\vecK_+\simeq\vecK_-$, and conversely, a $-4\pi$ vortex located at $\vecK_+$.
 If the second term in Eq.~\eqref{eq:dAvsdB} produces a wavefront dislocation in the LDOS, we can observe it directly in real space by filtering the Fourier components near $\Delta\vecK$, see Fig.~\ref{fig:filtered_LDOS}c,d, and Section~\ref{ssec:WRTB} for a detailed explanation of the filtering. We find that the low-energy theory (Eqs.~\eqref{eq:Green},\eqref{eq:LDOS}) matches these tight-binding results excellently (Fig.~\ref{fig:filtered_LDOS}e,f), providing the interpretation of the wavefront dislocation.

We next consider the general conditions, over a wide range of parameters, for observing the winding numbers in the LDOS.

\subsection{Winding region in the low-energy theory}
As discussed in detail in Ref.~\onlinecite{LenaCones}, the LDOS form in Eq.~\eqref{eq:dAvsdB} implies that on a circle of radius $r$ around the impurity there is a detectable wavefront dislocation (a difference $\Delta W$ in the number of wavefronts entering and exiting the circle) iff $\delta \rho_B(r,\omega) > \delta \rho_A(r,\omega)$.

\begin{figure}[h!]
\includegraphics[width=0.46\textwidth]{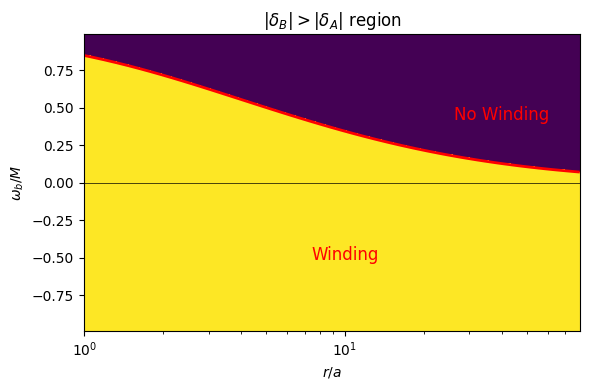}
\caption{Winding region in real space according to analytical low-energy theory of the hBN model. For a given impurity potential $V$ producing a bound state at energy $\omega_b$ (vertical axis), a circle of radius $r$ (horizontal axis) detects a wavefront dislocation (region "Winding" in plot) if the LDOS terms satisfy $\delta \rho_B(r,\omega_b)>\delta \rho_A(r,\omega_b)$ (see text). The "Winding" region corresponds to strong impurity potential (see Fig.~\ref{fig:wb_fV_all}).}
\label{fig:rhoA_rhoB}
\end{figure}

Figure~\ref{fig:rhoA_rhoB} reveals the real space region in which the LDOS of an impurity with potential $V$ exhibits a wavefront dislocation, in the case of the illustrative low-energy hBN model. For the given value of $V$ we find the bound state energy $\omega_b$ from Eq.~\eqref{eq:bs_linear_general} (see Fig.~\ref{fig:wb_fV_all}a), and then we find the winding region as the set of distances $r$ for which $\delta \rho_B(r,\omega_b) > \delta \rho_A(r,\omega_b)$.

Above a critical impurity strength $V_c$, the condition $\delta\rho_B(r,\omega_b)>\delta\rho_A(r,\omega_b)$ is satisfied for all distances $r$, such that the winding is, in principle, observable on any loop around the impurity. Clearly, the LDOS decays exponentially, so in STS data the signal would only be present in a finite region around the impurity. This result is in strong contrast to the gapless Dirac model of graphene, where the winding is observable only in some annuli around the impurity, not including a central disk centered on the impurity.

\subsection{Winding region in the tight-binding lattice model}
\label{ssec:WRTB}
We start by detailing the filtering procedure which was exemplified in Fig.~\ref{fig:filtered_LDOS}a,c,d. We first Fourier transform the LDOS image.

Second, a Gaussian filter is applied around the scattering feature of interest, here concretely the point $\vecK_+$. The width $\sigma$ of this filter must be chosen as a compromise: it should be narrow enough to isolate the contribution from the selected scattering vector, while remaining broad enough to retain information about varying spatial features. Third, we inverse Fourier transform the filtered data back into real space, and extract its complex phase $\phi(\vecr)$ (the data is complex because it contains only the data around $\vecK_+$, without the $\vecK_-$ counterpart). This results in an image akin to Fig.~\ref{fig:filtered_LDOS}d. Finally, we follow the pixels on a circle of given radius $r$, summing up the difference of phase values $\phi$ of neighboring pixels, which is a numerical implementation of $\int_0^{2\pi}\textrm{d}\theta\, \partial_\theta \phi=2\pi W_{\text{num}}$, giving the integer winding number $W_{\text{num}}$.

\begin{figure}[h!]
\includegraphics[width=0.46\textwidth]{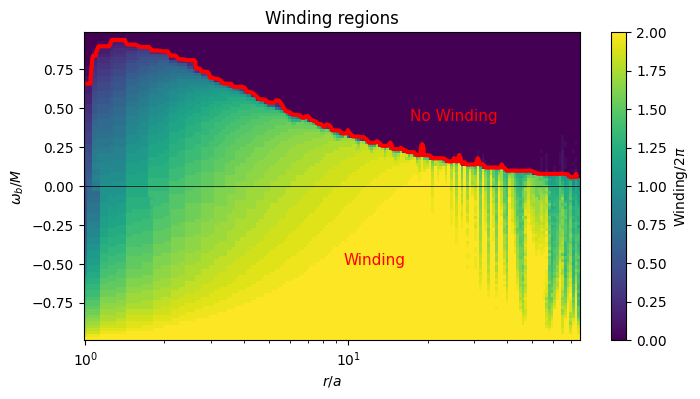}
\caption{Winding region in real space according to the tight-binding hBN model. For a given bound state at energy $\omega_b$ (vertical axis), we check if on a circle of radius $r$ (horizontal axis) we detect in the LDOS a wavefront dislocation (region "Winding" in plot). See text for the detection procedure.}
\label{fig:TB_rhoA_rhoB}
\end{figure}

Averaging the winding number $W_{\text{num}}$ for different values of $\sigma$ leads to the result in Fig.~\ref{fig:TB_rhoA_rhoB}, which is in very good agreement with the analytical low-energy prediction of Fig.~\ref{fig:rhoA_rhoB}. In the region marked as "Winding" the averaged $W_{\text{num}}$ is non-zero and matches well the value $\Delta W$.

As expected from the Fourier uncertainty principle, the choice of $\sigma$ affects the spatial resolution of the reconstructed phase $\phi(\vecr)$. Smaller values of $\sigma$ provide a more accurate description at large distances from the impurity, whereas larger values of $\sigma$ better capture the behavior in the immediate vicinity of the impurity. This trade-off is clearly illustrated in Appendix~\ref{sec:Filtering}, Fig.~\ref{fig:per_sigma}. In Fig.~\ref{fig:TB_rhoA_rhoB} we average over a range of $\sigma \in \left[0,\frac{|\Delta\vecK|}{2} \right]$ to analyze a very wide range of distances from the impurity, but any of the considered values of $\sigma$ gives the robust result already over a wide range of distances, from a few lattice constants to a hundred lattice constants.

\section{Berry curvature}
\label{sec:Berry curvature}
In this Section, we consider the Berry curvature\cite{Berry1984-gr,Xiao2010-tt,Xiao2012} as a possible observable in an STS experiment on an impurity. Gapless graphene has monopoles of Berry curvature at the $\vecK_\pm$ valleys, sourcing a quantized Berry phase on a loop surrounding the valley. In contrast, the opening of the gap in the hBN model makes the Berry curvature spread smoothly throughout the valley, and the Berry phase is not quantized, while the pseudospin winding number still is. We first establish that in fact the two can be related.

For a general two-band Bloch Hamiltonian, we first recall that the vector $\vec{h}$ (Eq.~\eqref{eq:vech}) provides curvature directly as:
\begin{equation}
    \Omega_\pm = \mp \frac{1}{2 |\vech|^3} \vech \cdot \left( \partial_x \vech \cross \partial_y \vech  \right),
\label{eq:2bandberry}
\end{equation}
where $\Omega_\pm$ is the Berry curvature in the upper/lower band, while we omit the explicit dependence on momentum for compactness. Our key step is to express $\Omega \equiv \Omega_+$ in polar coordinates $(q_x,q_y)=q(\cos(\theta_q),\sin(\theta_q))$ around a given point $\vecQ$, that will be a valley in the Brillouin zone. Writing $h_x + i h_y \equiv f e^{i\varphi}$:
\begin{multline}
\Omega(\vec {\vecQ} + \vecq) 
= - \frac{1}{2} \frac{h_z}{|\vech|^3} \frac{f}{q} \Bigl[
   \partial_\theta \varphi \left(\partial_q f - \frac{f}{h_z}  \partial_q h_z\right) \\
   - \partial_q \varphi \left( \partial_\theta f - \frac{f}{h_z}  \partial_\theta h_z\right)
   \Bigr]
\label{eq:polarberry}
\end{multline}

We now assume that the low-energy Hamiltonian around $\vecQ$ is
rotationally symmetric up to the phase of its off-diagonal component.
Accordingly, $f=f(q)$ and $h_z=h_z(q)$, whereas
\begin{equation*}
    \varphi(q,\theta_q)
    =
    W_Q\theta_q+\Phi(q,\theta_q),
\end{equation*}
where $W_Q$ is the winding number and $\Phi$ is single-valued, so that
\begin{equation*}
    \int_0^{2\pi}
    \frac{\mathrm{d}\theta_q}{2\pi}\,
    \partial_{\theta_q}\Phi
    =0.
\end{equation*}

We further assume that the gap at $\vecQ$ is carried by the diagonal
component and retain the leading terms of the low-energy expansion,
\begin{equation*}
    h_z(q)=\Delta_Q,
    \qquad
    f(q)=|\lambda|q^n,
\end{equation*}
where $n$ is a positive integer. Under these assumptions,
Eq.~\eqref{eq:polarberry} reduces, after angular averaging, to
\begin{equation}
    \widetilde{\Omega}_{\mathrm{cont}}(\vecQ+\vecq)
    =
    -\frac{nW_Q}{2}\,
    \frac{\Delta_Q|\lambda|^2|\vecq|^{2(n-1)}}
    {\left(
        \Delta_Q^2+|\lambda|^2|\vecq|^{2n}
    \right)^{3/2}} .
    \label{eq:continuumberry}
\end{equation}
We refer to Eq.~\eqref{eq:continuumberry} as the continuum-theory
prediction. It retains the full momentum dependence of the
low-energy Hamiltonian within the valley approximation.

Expanding Eq.~\eqref{eq:continuumberry} at small momentum gives the
leading-order prediction
\begin{equation}
    \widetilde{\Omega}_{\mathrm{LO}}(\vecQ+\vecq)
    =
    -\frac{nW_Q}{2}\,
    \operatorname{sgn}(\Delta_Q)
    \left(\frac{|\lambda|}{\Delta_Q}\right)^{\!2}
    |\vecq|^{2(n-1)} .
    \label{eq:Taylorberry}
\end{equation}
Equation~\eqref{eq:Taylorberry} is valid only in the immediate
vicinity of $\vecQ$, whereas Eq.~\eqref{eq:continuumberry} describes
the momentum dependence over the range in which the continuum
Hamiltonian remains applicable.

The formulas Eq.~\eqref{eq:polarberry} and Eq.~\eqref{eq:Taylorberry} are the central result of this Section, showing that for quite generic isotropic dispersions in two-band models, the value of the Berry curvature in a valley can be extracted (up to a sign, and at least to some low order in $q$) from the winding number and the spectroscopic parameters of the energy dispersion. As we have shown in previous sections, the winding number itself can be in principle obtained from STS on impurities. Conversely, the result shows that the observation of a winding number implies a non trivial Berry curvature.

\subsection{Application to isotropic valley models}
\label{ssec:Berry curvature isotropic models}

\begin{figure}[h!]
\includegraphics[width=0.49\textwidth]{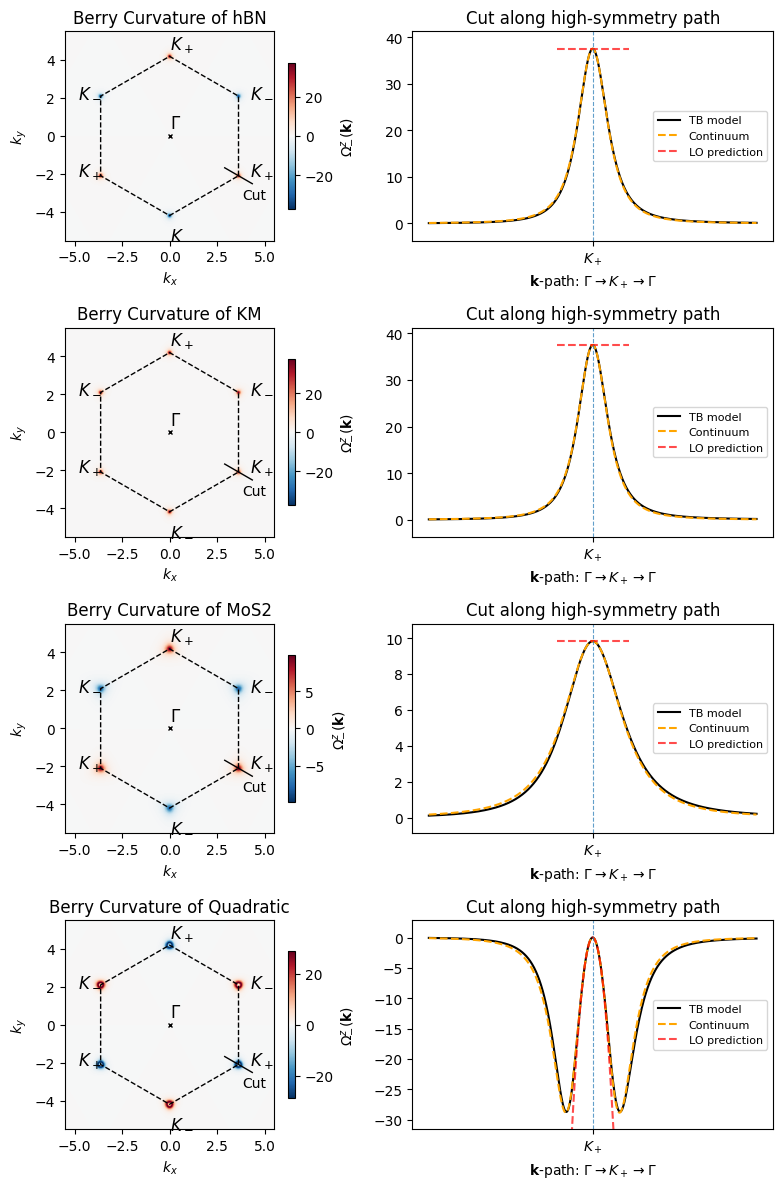}
\caption{Berry curvature for the hBN, Kane--Mele, MoS$_2$, and quadratic models (from top to bottom). Left column: tight-binding Berry curvature over the first Brillouin zone. Right column: cut through the $\vecK_+$ valley along the indicated high-symmetry path. Tight-binding results (black solid lines) are compared with the continuum expressions (orange dashed lines) and with the leading-order expansion of Eq.~\eqref{eq:Taylorberry} (red dashed lines). The linear Dirac models have a finite Berry curvature at the valley centre, whose magnitude is correctly predicted by the leading-order expansion. For the quadratic model, the Berry curvature vanishes at $\vecK_+$ and grows quadratically away from it, producing an annular distribution. Momenta are expressed in
units of $a^{-1}$.}
\label{fig:BC_comparison}
\end{figure}

We now apply Eq.~\eqref{eq:Taylorberry} to four isotropic valley
models: hBN, the pure Kane--Mele model, the TMD
monolayer MoS$_2$, and the quadratic band model. The results are
summarized in Fig.~\ref{fig:BC_comparison}. In each row, the
tight-binding Berry curvature over the Brillouin zone is shown
together with a cut through the $\vecK_+$ valley. The latter is
compared with both the corresponding continuum theory and the
leading-order prediction obtained from Eq.~\eqref{eq:Taylorberry}.

\subsubsection{Linear Dirac models.}

\paragraph{} For the hBN model introduced in Eq.~\eqref{eq:linearized_hBN}, the
off-diagonal component is linear in momentum, so that $n=1$, while
the winding at valley $\vecK_\xi$ is $W_{K_\xi}=-\xi$ in our
convention. Equation~\eqref{eq:Taylorberry} therefore predicts a
finite Berry curvature at the valley centre,
\begin{equation}
    \widetilde{\Omega}(\vecK_\xi)
    =
    \frac{\xi}{2}
    \frac{\operatorname{sgn}(M)v_F^2}{M^2},
\end{equation}
up to the band and Hamiltonian sign conventions used above. The
opposite windings of the two valleys consequently produce Berry
curvatures of opposite sign. As shown in the first row of
Fig.~\ref{fig:BC_comparison}, the predicted valley-centre value agrees
with both the continuum and tight-binding calculations.

\paragraph{} The pure Kane--Mele model provides the corresponding spin-dependent
benchmark. For a fixed spin sector, the intrinsic spin--orbit term
acts as a valley-dependent Dirac mass. Its change of sign between the
two valleys compensates the change in the valley winding, so that the
Berry curvature has the same sign at $\vecK_+$ and $\vecK_-$ for a
given spin. The second row of Fig.~\ref{fig:BC_comparison} again shows
that Eq.~\eqref{eq:Taylorberry} correctly reproduces the value of the
Berry curvature at the valley centre.

\paragraph{} A more experimentally relevant SOC example is provided by monolayer
MoS$_2$, for which the Semenoff, Kane--Mele, and valley--Zeeman terms
are simultaneously present. These terms modify the effective gap in
each spin and valley sector but do not change the linear character of
the dispersion. Consequently, $n=1$ and the Berry curvature remains
finite at the valley centre. Using the appropriate sector-dependent
gap in Eq.~\eqref{eq:Taylorberry} gives the leading-order prediction
shown in the third row of Fig.~\ref{fig:BC_comparison}, in agreement
with the continuum and tight-binding results.

\subsubsection{Quadratic band touching.}

The quadratic model of Eq.~\eqref{eq:linearized_quad} has
$f(q)=|\beta|q^2$, and hence $n=2$, together with the doubled winding
$W_{K_\xi}=2\xi$. Equation~\eqref{eq:Taylorberry} then gives
\begin{equation}
    \widetilde{\Omega}(\vecK_\xi+\vecq)
    =
    -2\xi\,\operatorname{sgn}(M)
    \frac{|\beta|^2}{M^2}q^2
    +O(q^4).
\end{equation}
Unlike in the linear Dirac models, the Berry curvature therefore
vanishes at the valley centre and grows quadratically away from it.
This accounts for the annular distributions seen in the last row of
Fig.~\ref{fig:BC_comparison}. The leading-order expression captures
the behaviour close to $\vecK_\xi$, while the full continuum theory
is required to reproduce the position and magnitude of the extrema
at finite momentum.

The comparison in Fig.~\ref{fig:BC_comparison} illustrates the two
pieces of information contained in Eq.~\eqref{eq:Taylorberry}. The
winding number determines the sign and angular-topological content of
the valley contribution, whereas the dispersion exponent determines
the radial onset of the Berry curvature: it is finite at the valley
centre for $n=1$ and vanishes as $q^2$ for $n=2$. Thus, within the
isotropic two-band setting considered here, impurity spectroscopy can
constrain the local Berry curvature through the winding extracted
from the LDOS and the independently measurable low-energy dispersion.

This correspondence is not general beyond the assumptions entering
Eq.~\eqref{eq:Taylorberry}. In particular, anisotropic valleys cannot
in general be characterized by the radial average used above. We
return to this point for the hybrid and semi-Dirac models in
Section~\ref{sec:Hybrid}.

\subsection{The anisotropic semi-Dirac model}
\label{sec:Hybrid}
\begin{figure}[h!]
\includegraphics[width=0.5\textwidth]{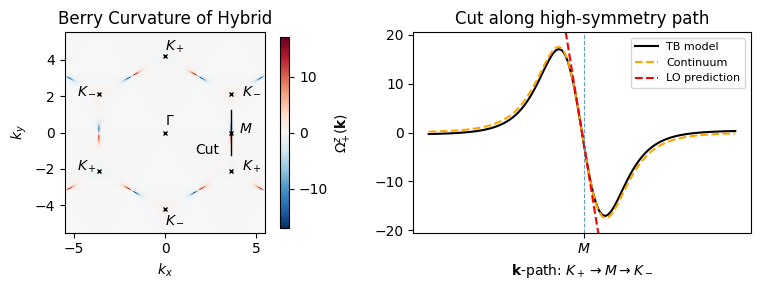}
\caption{Berry curvature of the semi-Dirac model [$M=19$, $t=40/3$, $t_2=10$]. We show the whole Brillouin zone (on the left) and a cut along a line at the $\vecM$-valley (on the right). This system has a gap opening at the three $\vecM$-points of size $1$. The units of the axes are given in terms of $a^{-1}$. The curvature presents a dipole form, predicted analytically eq.~\eqref{eq:dipole_hybrid}.}
\label{fig:BC_hybrid}
\end{figure}

Here we discuss a more complicated anisotropic valley dispersion, which is exemplified by the semi-Dirac model. By tuning the parameters of Eq.~\eqref{eq:H0_NNNN} as follows: $t_1 = 3t_3 = 3t$ and $M = 2 t_2 + \delta$, where $\delta$ is a small detuning that we call the gapping parameter, we obtain three valleys at the three $\vecM$-points on the edges of the Brillouin zone. The Hamiltonian in each valley is anisotropic, and in the particular example of the $\vecM$-point marked in Fig.~\ref{fig:BC_hybrid}:
\begin{equation}
    H_0^{\textrm{semi}}(\vecM+\vecq) = 
\begin{pmatrix}
\delta + C q_y^2 & -i v_x^* q_x + D^* q_y^2 \\
i v_x q_x + D q_y^2   & -\delta - C q_y^2
\end{pmatrix},
\label{eq:linearised_H_hybrid}
\end{equation}
with $v_x = e^{i\pi/3} 2\sqrt{3}t$, $C\equiv-\frac{t_2}{2}$, and $D\equiv e^{i\pi/3} \frac{t}{4}$. As detailed in Appendix~\ref{app:Hybrid}, the low-energy dispersion is semi-Dirac, meaning linear in one direction and quadratic in the other:
\begin{equation}
\epsilon(\vecM+\vecq) \approx \delta +  |v_x| |q_x| + \beta q_y^2,
\end{equation}
in the range of momenta $\delta/t\ll|\vecq|/|\vecM|\ll1$, and where we introduced $\beta = \sqrt{C^2 + |D|^2}$.

The semi-Dirac valleys at the $\vecM$-points arise because the tuning of the hoppings merges the two Dirac cones from the original $\vecK_\pm$ points, which have opposite windings. Hence, the pseudospin winding number $W_M=0$ in these $\vecM$ semi-Dirac valleys. This can be confirmed by writing the off-diagonal term of the Hamiltonian Eq.~\eqref{eq:linearised_H_hybrid} in polar coordinates, concretely, since 
\begin{align}
    \varphi(\vecM+\vecq) - \frac{\pi}3 = \textrm{Arg}(i |v_x| q_x + |D| q_y^2) \in \left[-\frac{\pi}2, \frac{\pi}2 \right],
\end{align}
the phase cannot achieve any winding around the valley as $\theta_q$ goes from $0$ to $2\pi$.

The continuum theory result for the Berry curvature along the $y$-axis of the valley Hamiltonian Eq.~\eqref{eq:linearised_H_hybrid} follows as:
\begin{equation}
\Omega_\pm(\mathbf M + q_y\mathbf{e}_y) = \pm \textrm{sgn}(\delta) \frac{|v_xD|}{\delta^2 } \frac{q_y}{\left[1+ \frac{2C}{\delta} q_y^2 + \frac{(C^2 + |D|^2)}{\delta^2}q_y^4\right]^{3/2}}.
\end{equation}
To lowest (linear) order we get:
\begin{equation}
    \Omega_\pm(\vecM + q_y\mathbf{e}_y)
    \simeq
    \pm \textrm{sgn}(\delta)
    \frac{|v_x D|}{\delta^2}\,q_y,
\label{eq:dipole_hybrid}
\end{equation}
showing that the Berry curvature has a dipole form, changing sign as $q_y$ changes sign.

This dipole-like profile of the Berry curvature at the $M$-valley leads to the vanishing of the averaged curvature $\tilde{\Omega}$, consistent with Eq.~\eqref{eq:polarberry} due to the fact that the winding number is zero.

We may bring out the essence of this connection between the vanishing of the winding number and of the averaged Berry curvature by considering a simpler toy model:
\begin{equation}
H^{\textrm{toy}}(\vecq)=
\begin{pmatrix}
M & v q e^{i|\theta_q - \theta_0|} \\
v q e^{-i|\theta_q - \theta_0|} & -M
\end{pmatrix},
\end{equation}
which produces an isotropic dispersion with a vanishing winding number $W=0$ and a non-zero Berry curvature. Namely, to lowest order in $q$, the Berry curvature without averaging, Eq.~\eqref{eq:polarberry}, simplifies to
\begin{equation}
\Omega_\pm(\vecq) \simeq \pm \frac{1}{2} \frac{\sign(M)}{M^2} |v|^2 \textrm{sgn}(\theta_q-\theta_0),
\end{equation}
exhibiting a clear dipolar form with opposite signs on two halves of the $\vecq$-plane. Again, the averaging of the Berry curvature on a circle around the origin will lead to zero, consistent with the vanishing of the winding number.

\section{Conclusions}
\label{sec:conclusions}
In this work, we studied impurity-induced bound states and quasiparticle interference in a broad class of gapped Dirac and semi-Dirac systems. Using both tight-binding simulations and analytical Green’s function methods, we showed that the Fourier-filtered LDOS around impurities exhibits quantized phase windings governed by the winding number of the pseudo-spin in the Bloch Hamiltonian.

For gapped Dirac systems, such as relevant for hBN and monolayer TMDs, the impurity-bound-state LDOS displays characteristic dislocations that can, in principle, be observed in STM experiments. In fact, for strong enough impurity potentials, we expect a clear dislocation pattern in the LDOS at any distance from the impurity.

We further clarified the relation between pseudo-spin winding and Berry curvature. A non-trivial winding implies a finite Berry curvature in the valley, and we show the striking fact that the momentum-dependent Berry curvature can be directly deduced using the winding number and the shape of the bulk dispersion, at least for dispersions that are isotropic in the valley. On the example of the anisotropic semi-Dirac system, we also show that the winding can be related to a dipolar structure of the Berry curvature.

Our results therefore establish impurity-induced quasiparticle interference as a sensitive real-space probe of wave-function geometry and topology in gapped Dirac materials.

\begin{acknowledgments}
We are grateful to M.-O. Goerbig, F. Piéchon, G. Montambaux and J.-N. Fuchs for insightful discussions. We would also like to acknowledge the support of the French Agence Nationale de la Recherche (ANR), under grant number ANR-22-CE300037.
\end{acknowledgments}

\bibliography{main.bib}

@article{McCann12,
  title = {$z\ensuremath{\rightarrow}\ensuremath{-}z$ Symmetry of Spin-Orbit Coupling and Weak Localization in Graphene},
  author = {McCann, Edward and Fal'ko, Vladimir I.},
  journal = {Phys. Rev. Lett.},
  volume = {108},
  issue = {16},
  pages = {166606},
  numpages = {5},
  year = {2012},
  month = {Apr},
  publisher = {American Physical Society},
  doi = {10.1103/PhysRevLett.108.166606},
  url = {https://link.aps.org/doi/10.1103/PhysRevLett.108.166606}
}

@article{Wang16,
  title = {Origin and Magnitude of `Designer' Spin-Orbit Interaction in Graphene on Semiconducting Transition Metal Dichalcogenides},
  author = {Wang, Zhe and Ki, Dong-Keun and Khoo, Jun Yong and Mauro, Diego and Berger, Helmuth and Levitov, Leonid S. and Morpurgo, Alberto F.},
  journal = {Phys. Rev. X},
  volume = {6},
  issue = {4},
  pages = {041020},
  numpages = {15},
  year = {2016},
  month = {Oct},
  publisher = {American Physical Society},
  doi = {10.1103/PhysRevX.6.041020},
  url = {https://link.aps.org/doi/10.1103/PhysRevX.6.041020}
}

@article{Fabian17,
  title = {Model spin-orbit coupling Hamiltonians for graphene systems},
  author = {Kochan, Denis and Irmer, Susanne and Fabian, Jaroslav},
  journal = {Phys. Rev. B},
  volume = {95},
  issue = {16},
  pages = {165415},
  numpages = {19},
  year = {2017},
  month = {Apr},
  publisher = {American Physical Society},
  doi = {10.1103/PhysRevB.95.165415},
  url = {https://link.aps.org/doi/10.1103/PhysRevB.95.165415}
}

@article{Shiba1968,
    author = {Shiba, Hiroyuki},
    title = {Classical Spins in Superconductors},
    journal = {Progress of Theoretical Physics},
    volume = {40},
    number = {3},
    pages = {435-451},
    year = {1968},
    month = {09},
    issn = {0033-068X},
    doi = {10.1143/PTP.40.435},
    url = {https://doi.org/10.1143/PTP.40.435},
    eprint = {https://academic.oup.com/ptp/article-pdf/40/3/435/5185550/40-3-435.pdf},
}

@article{Wehling2009,
title = {Adsorbates on graphene: Impurity states and electron scattering},
journal = {Chemical Physics Letters},
volume = {476},
number = {4},
pages = {125-134},
year = {2009},
issn = {0009-2614},
doi = {https://doi.org/10.1016/j.cplett.2009.06.005},
url = {https://www.sciencedirect.com/science/article/pii/S0009261409006642},
author = {T.O. Wehling and M.I. Katsnelson and A.I. Lichtenstein}
}

@article{black2015,
  title={Filling of magnetic-impurity-induced gap in topological insulators by potential scattering},
  author={Black-Schaffer, Annica M and Balatsky, AV and Fransson, Jonas},
  journal={Physical Review B},
  volume={91},
  number={20},
  pages={201411},
  year={2015},
  publisher={APS}
}

@article{LenaCones,
  title = {Detecting the topological winding of superconducting nodes via local density of states},
  author = {Engstr\"om, Lena and Simon, Pascal and Mesaros, Andrej},
  journal = {Phys. Rev. B},
  volume = {111},
  issue = {13},
  pages = {134505},
  numpages = {17},
  year = {2025},
  month = {Apr},
  publisher = {American Physical Society},
  doi = {10.1103/PhysRevB.111.134505},
  url = {https://link.aps.org/doi/10.1103/PhysRevB.111.134505}
}

@misc{Akkermans2025,
  title         = "Topological winding numbers from wavefront dislocations in
                   local electronic density",
  author        = "Abulafia, Yuval and Akkermans, Eric",
  month         =  aug,
  year          =  2025,
  copyright     = "http://arxiv.org/licenses/nonexclusive-distrib/1.0/",
  archivePrefix = "arXiv",
  primaryClass  = "cond-mat.mes-hall",
  eprint        = "2508.19128"
}

@misc{Akkermans2023,
  title         = "Wavefronts dislocations measure topology in graphene with
                   defects",
  author        = "Abulafia, Yuval and Goft, Amit and Orion, Nadav and
                   Akkermans, Eric",
  month         =  jul,
  year          =  2023,
  copyright     = "http://arxiv.org/licenses/nonexclusive-distrib/1.0/",
  archivePrefix = "arXiv",
  primaryClass  = "cond-mat.mes-hall",
  eprint        = "2307.05185"
}

@ARTICLE{Aivazian2015,
  title     = "Magnetic control of valley pseudospin in monolayer {WSe2}",
  author    = "Aivazian, G and Gong, Zhirui and Jones, Aaron M and Chu, Rui-Lin
               and Yan, J and Mandrus, D G and Zhang, Chuanwei and Cobden,
               David and Yao, Wang and Xu, X",
  journal   = "Nat. Phys.",
  publisher = "Springer Science and Business Media LLC",
  volume    =  11,
  number    =  2,
  pages     = "148--152",
  month     =  feb,
  year      =  2015,
  language  = "en"
}

@ARTICLE{Wang2017,
  title     = "Valley- and spin-polarized Landau levels in monolayer {WSe2}",
  author    = "Wang, Zefang and Shan, Jie and Mak, Kin Fai",
  journal   = "Nat. Nanotechnol.",
  publisher = "Springer Science and Business Media LLC",
  volume    =  12,
  number    =  2,
  pages     = "144--149",
  month     =  feb,
  year      =  2017,
  language  = "en"
}

@ARTICLE{Xiao2012,
  title     = "Coupled spin and valley physics in monolayers of {MoS2} and
               other {group-VI} dichalcogenides",
  author    = "Xiao, Di and Liu, Gui-Bin and Feng, Wanxiang and Xu, Xiaodong
               and Yao, Wang",
  journal   = "Phys. Rev. Lett.",
  publisher = "American Physical Society (APS)",
  volume    =  108,
  number    =  19,
  pages     = "196802",
  month     =  may,
  year      =  2012,
  copyright = "http://link.aps.org/licenses/aps-default-license",
  language  = "en"
}

@ARTICLE{Bieniek2018,
  title     = "Band nesting, massive Dirac fermions, and valley Land{\'e} and
               Zeeman effects in transition metal dichalcogenides: A
               tight-binding model",
  author    = "Bieniek, Maciej and Korkusi{\'n}ski, Marek and Szulakowska,
               Ludmi{\l}a and Potasz, Pawe{\l} and Ozfidan, Isil and Hawrylak,
               Pawe{\l}",
  journal   = "Phys. Rev. B.",
  publisher = "American Physical Society (APS)",
  volume    =  97,
  number    =  8,
  month     =  feb,
  year      =  2018,
  copyright = "https://link.aps.org/licenses/aps-default-license"
}

@ARTICLE{Sheina2023,
  title     = "Hydrogenic spin-valley states of the bromine donor in {2H-MoTe2}",
  author    = "Sheina, Valeria and Lang, Guillaume and Stolyarov, Vasily and
               Marchenkov, Vyacheslav and Naumov, Sergey and Perevalova,
               Alexandra and Girard, Jean-Christophe and Rodary, Guillemin and
               David, Christophe and Sop, Leonnel Romuald and Pierucci, Debora
               and Ouerghi, Abdelkarim and Cantin, Jean-Louis and Leridon,
               Brigitte and Ghorbani-Asl, Mahdi and Krasheninnikov, Arkady V
               and Aubin, Herv{\'e}",
  journal   = "Commun. Phys.",
  publisher = "Springer Science and Business Media LLC",
  volume    =  6,
  number    =  1,
  month     =  jun,
  year      =  2023,
  copyright = "https://creativecommons.org/licenses/by/4.0",
  language  = "en"
}

@ARTICLE{Park2011,
  title     = "Berry phase and pseudospin winding number in bilayer graphene",
  author    = "Park, Cheol-Hwan and Marzari, Nicola",
  journal   = "Phys. Rev. B Condens. Matter Mater. Phys.",
  publisher = "American Physical Society (APS)",
  volume    =  84,
  number    =  20,
  month     =  nov,
  year      =  2011,
  copyright = "http://link.aps.org/licenses/aps-default-license"
}

@ARTICLE{Novoselov2006,
  title     = "Unconventional quantum Hall effect and Berry's phase of 2$\pi$
               in bilayer graphene",
  author    = "Novoselov, K S and McCann, E and Morozov, S V and Fal'ko, V I
               and Katsnelson, M I and Zeitler, U and Jiang, D and Schedin, F
               and Geim, A K",
  journal   = "Nat. Phys.",
  publisher = "Springer Science and Business Media LLC",
  volume    =  2,
  number    =  3,
  pages     = "177--180",
  month     =  mar,
  year      =  2006,
  language  = "en"
}

@article{Chen2017,
doi = {10.1088/1361-648X/aa54da},
url = {https://dx.doi.org/10.1088/1361-648X/aa54da},
year = {2017},
month = {jan},
publisher = {IOP Publishing},
volume = {29},
number = {10},
pages = {103001},
author = {Chen, Lan and Cheng, Peng and Wu, Kehui},
title = {{Quasiparticle interference in unconventional 2D systems}},
journal = {Journal of Physics: Condensed Matter}
}

@misc{Chi2017,
  title         = "Determination of the superconducting order parameter from defect bound state quasiparticle interference",
  author        = "Chi, Shun and Hardy, W N and Liang, Ruixing and Dosanjh, P and Wahl, Peter and Burke, S A and Bonn, D A",
  month         =  oct,
  year          =  2017,
  archivePrefix = "arXiv",
  primaryClass  = "cond-mat.supr-con",
  eprint        = "1710.09089"
}

@ARTICLE{Bhattacharyya2023,
  title     = "Superconducting gap symmetry from Bogoliubov quasiparticle
               interference analysis on {Sr2RuO4}",
  author    = "Bhattacharyya, Shinibali and Kreisel, Andreas and Kong, X and
               Berlijn, T and R{\o}mer, Astrid T and Andersen, Brian M and
               Hirschfeld, P J",
  journal   = "Phys. Rev. B.",
  publisher = "American Physical Society (APS)",
  volume    =  107,
  number    =  14,
  month     =  apr,
  year      =  2023,
  copyright = "https://link.aps.org/licenses/aps-default-license",
  language  = "en"
}

@article{Nunner2006,
  title = {{Fourier transform spectroscopy of $d$-wave quasiparticles in the presence of atomic scale pairing disorder}},
  author = {Nunner, Tamara S. and Chen, Wei and Andersen, Brian M. and Melikyan, Ashot and Hirschfeld, P. J.},
  journal = {Phys. Rev. B},
  volume = {73},
  issue = {10},
  pages = {104511},
  numpages = {7},
  year = {2006},
  month = {Mar},
  publisher = {American Physical Society},
  doi = {10.1103/PhysRevB.73.104511},
  url = {https://link.aps.org/doi/10.1103/PhysRevB.73.104511}
}

@article{Pereg-Barnea2008,
  title = {{Magnetic-field dependence of quasiparticle interference peaks in a $d$-wave superconductor with weak disorder}},
  author = {Pereg-Barnea, T. and Franz, M.},
  journal = {Phys. Rev. B},
  volume = {78},
  issue = {2},
  pages = {020509},
  numpages = {4},
  year = {2008},
  month = {Jul},
  publisher = {American Physical Society},
  doi = {10.1103/PhysRevB.78.020509},
  url = {https://link.aps.org/doi/10.1103/PhysRevB.78.020509}
}

@article{Hirschfeld2021,
  title = {Robust determination of the superconducting gap sign structure via quasiparticle interference},
  author = {Hirschfeld, P. J. and Altenfeld, D. and Eremin, I. and Mazin, I. I.},
  journal = {Phys. Rev. B},
  volume = {92},
  issue = {18},
  pages = {184513},
  numpages = {13},
  year = {2015},
  month = {Nov},
  publisher = {American Physical Society},
  doi = {10.1103/PhysRevB.92.184513},
  url = {https://link.aps.org/doi/10.1103/PhysRevB.92.184513}
}

@article{SeamusDwave,
  title = {Phase-sensitive determination of nodal $d$-wave order parameter in single-band and multiband superconductors},
  author = {B\"oker, Jakob and Sulangi, Miguel Antonio and Akbari, Alireza and Davis, J. C. S\'eamus and Hirschfeld, P. J. and Eremin, Ilya M.},
  journal = {Phys. Rev. B},
  volume = {101},
  issue = {21},
  pages = {214505},
  numpages = {11},
  year = {2020},
  month = {Jun},
  publisher = {American Physical Society},
  doi = {10.1103/PhysRevB.101.214505},
  url = {https://link.aps.org/doi/10.1103/PhysRevB.101.214505}
}

@article{SeamusFwave,
  title={Imaging Cooper pairing of heavy fermions in CeCoIn5},
  author={Allan, Milan Peter and Massee, F and Morr, DK and Van Dyke, J and Rost, Andreas Winfried and Mackenzie, AP and Petrovic, C and Davis, James C},
  journal={Nature physics},
  volume={9},
  number={8},
  pages={468--473},
  year={2013},
  publisher={Nature Publishing Group UK London}
}

@ARTICLE{Hoffman2002,
  title     = "Imaging quasiparticle interference in {Bi2Sr2CaCu2O8+delta}",
  author    = "Hoffman, J E and McElroy, K and Lee, D-H and Lang, K M and
               Eisaki, H and Uchida, S and Davis, J C",
  journal   = "Science",
  publisher = "American Association for the Advancement of Science (AAAS)",
  volume    =  297,
  number    =  5584,
  pages     = "1148--1151",
  month     =  aug,
  year      =  2002,
  language  = "en"
}

@ARTICLE{Wang2003,
  title     = "Quasiparticle scattering interference in high-temperature
               superconductors",
  author    = "Wang, Qiang-Hua and Lee, Dung-Hai",
  journal   = "Phys. Rev. B Condens. Matter",
  publisher = "American Physical Society (APS)",
  volume    =  67,
  number    =  2,
  month     =  jan,
  year      =  2003,
  copyright = "http://link.aps.org/licenses/aps-default-license"
}

@ARTICLE{Balatsky2006,
  title     = "Impurity-induced states in conventional and unconventional
               superconductors",
  author    = "Balatsky, A V and Vekhter, I and Zhu, Jian-Xin",
  journal   = "Rev. Mod. Phys.",
  publisher = "American Physical Society (APS)",
  volume    =  78,
  number    =  2,
  pages     = "373--433",
  month     =  may,
  year      =  2006,
  copyright = "http://link.aps.org/licenses/aps-default-license",
  language  = "en"
}

@article{Dutreix2019,
  author    = {Dutreix, C and Gonz{\'a}lez-Herrero, H and Brihuega, I and others},
  title     = {Measuring the Berry phase of graphene from wavefront dislocations in Friedel oscillations},
  journal   = {Nature},
  year      = {2019},
  volume    = {574},
  pages     = {219--222},
  doi       = {10.1038/s41586-019-1613-5},
  publisher = {Springer Nature},
  month     = oct
}

@ARTICLE{Pereg-BarneaMacdonald2008,
  title     = "Chiral quasiparticle local density of states maps in graphene",
  author    = "Pereg-Barnea, T and MacDonald, A H",
  journal   = "Phys. Rev. B Condens. Matter Mater. Phys.",
  publisher = "American Physical Society (APS)",
  volume    =  78,
  number    =  1,
  month     =  jul,
  year      =  2008,
  copyright = "http://link.aps.org/licenses/aps-default-license",
  language  = "en"
}

@article{Wehling2007,
  title = {Local electronic signatures of impurity states in graphene},
  author = {Wehling, T. O. and Balatsky, A. V. and Katsnelson, M. I. and Lichtenstein, A. I. and Scharnberg, K. and Wiesendanger, R.},
  journal = {Phys. Rev. B},
  volume = {75},
  issue = {12},
  pages = {125425},
  numpages = {5},
  year = {2007},
  month = {Mar},
  publisher = {American Physical Society},
  doi = {10.1103/PhysRevB.75.125425},
  url = {https://link.aps.org/doi/10.1103/PhysRevB.75.125425}
}

@ARTICLE{Castro_Neto2009-wu,
  title     = "The electronic properties of graphene",
  author    = "Castro Neto, A H and Guinea, F and Peres, N M R and Novoselov, K
               S and Geim, A K",
  journal   = "Rev. Mod. Phys.",
  publisher = "American Physical Society (APS)",
  volume    =  81,
  number    =  1,
  pages     = "109--162",
  month     =  jan,
  year      =  2009,
  copyright = "http://link.aps.org/licenses/aps-default-license",
  language  = "en"
}

@ARTICLE{Roushan2009-ge,
  title     = "Topological surface states protected from backscattering by
               chiral spin texture",
  author    = "Roushan, Pedram and Seo, Jungpil and Parker, Colin V and Hor, Y
               S and Hsieh, D and Qian, Dong and Richardella, Anthony and
               Hasan, M Z and Cava, R J and Yazdani, Ali",
  journal   = "Nature",
  publisher = "Springer Science and Business Media LLC",
  volume    =  460,
  number    =  7259,
  pages     = "1106--1109",
  month     =  aug,
  year      =  2009,
  language  = "en"
}

@ARTICLE{Xu2014-go,
  title     = "Spin and pseudospins in layered transition metal dichalcogenides",
  author    = "Xu, Xiaodong and Yao, Wang and Xiao, Di and Heinz, Tony F",
  journal   = "Nat. Phys.",
  publisher = "Springer Science and Business Media LLC",
  volume    =  10,
  number    =  5,
  pages     = "343--350",
  month     =  may,
  year      =  2014,
  language  = "en"
}

@ARTICLE{Bena2008-hs,
  title     = "Effect of a single localized impurity on the local density of
               States in monolayer and bilayer graphene",
  author    = "Bena, Cristina",
  journal   = "Phys. Rev. Lett.",
  publisher = "American Physical Society (APS)",
  volume    =  100,
  number    =  7,
  pages     = "076601",
  month     =  feb,
  year      =  2008,
  copyright = "http://link.aps.org/licenses/aps-default-license",
  language  = "en"
}

@ARTICLE{Semenoff1984-od,
  title     = "Condensed-matter simulation of a three-dimensional anomaly",
  author    = "Semenoff, Gordon W",
  journal   = "Phys. Rev. Lett.",
  publisher = "American Physical Society (APS)",
  volume    =  53,
  number    =  26,
  pages     = "2449--2452",
  month     =  dec,
  year      =  1984,
  copyright = "http://link.aps.org/licenses/aps-default-license",
  language  = "en"
}

@ARTICLE{Kane2005-cf,
  title     = "Quantum spin Hall effect in graphene",
  author    = "Kane, C L and Mele, E J",
  journal   = "Phys. Rev. Lett.",
  publisher = "American Physical Society (APS)",
  volume    =  95,
  number    =  22,
  pages     = "226801",
  month     =  nov,
  year      =  2005,
  copyright = "http://link.aps.org/licenses/aps-default-license",
  language  = "en"
}

@ARTICLE{Montambaux2009-yn,
  title     = "Merging of Dirac points in a two-dimensional crystal",
  author    = "Montambaux, G and Pi{\'e}chon, F and Fuchs, J-N and Goerbig, M O",
  journal   = "Phys. Rev. B Condens. Matter Mater. Phys.",
  publisher = "American Physical Society (APS)",
  volume    =  80,
  number    =  15,
  month     =  oct,
  year      =  2009,
  copyright = "http://link.aps.org/licenses/aps-default-license"
}

@ARTICLE{Montambaux2009-nv,
  title     = "A universal Hamiltonian for motion and merging of Dirac points
               in a two-dimensional crystal",
  author    = "Montambaux, G and Pi{\'e}chon, F and Fuchs, J-N and Goerbig, M O",
  journal   = "Eur. Phys. J. B",
  publisher = "Springer Science and Business Media LLC",
  volume    =  72,
  number    =  4,
  pages     = "509--520",
  month     =  dec,
  year      =  2009,
  language  = "en"
}

@ARTICLE{De_Gail2012-pk,
  title     = "Manipulation of Dirac points in graphene-like crystals",
  author    = "de Gail, R and Fuchs, J-N and Goerbig, M O and Pi{\'e}chon, F
               and Montambaux, G",
  journal   = "Physica B Condens. Matter",
  publisher = "Elsevier BV",
  volume    =  407,
  number    =  11,
  pages     = "1948--1952",
  month     =  jun,
  year      =  2012,
  language  = "en"
}

@ARTICLE{Berry1984-gr,
  title     = "Quantal phase factors accompanying adiabatic changes",
  author    = "Berry, Michael Victor",
  journal   = "Proc. R. Soc. Lond.",
  publisher = "The Royal Society",
  volume    =  392,
  number    =  1802,
  pages     = "45--57",
  month     =  mar,
  year      =  1984,
  language  = "en"
}

@ARTICLE{Xiao2010-tt,
  title     = "Berry phase effects on electronic properties",
  author    = "Xiao, Di and Chang, Ming-Che and Niu, Qian",
  journal   = "Rev. Mod. Phys.",
  publisher = "American Physical Society (APS)",
  volume    =  82,
  number    =  3,
  pages     = "1959--2007",
  month     =  jul,
  year      =  2010,
  copyright = "http://link.aps.org/licenses/aps-default-license",
  language  = "en"
}

\appendix
\begin{widetext}

\section{Low-energy expansions of the lattice structure factors}
\label{sec:structure_factor_expansions}

The continuum Hamiltonians used throughout the main text are obtained by expanding the lattice structure factors around the relevant high-symmetry points of the Brillouin zone. We collect here the low-energy expansions of the nearest-, second-, and third-neighbor structure factors, denoted by $f_1$, $f_2$, and $f_3$, together with the spin-orbit structure factor $F$. The expansions are performed around the valleys
\begin{equation}
\mathbf K^\xi
=
\frac{2\pi}{\sqrt3}\mathbf e_x
-
\xi\frac{2\pi}{3}\mathbf e_y,
\qquad
\xi=\pm1,
\end{equation}
and around the $M$ point
\begin{equation}
\mathbf M=\frac{2\pi}{\sqrt3}\mathbf e_x.
\end{equation}
We use the primitive lattice vectors
\begin{equation}
\mathbf a_1=
\frac{\sqrt3}{2}\mathbf e_x-\frac12\mathbf e_y,
\qquad
\mathbf a_2=
\frac{\sqrt3}{2}\mathbf e_x+\frac12\mathbf e_y.
\end{equation}
Throughout this appendix, $\mathbf q=(q_x,q_y)$ denotes a small momentum measured from either $\mathbf K^\xi$ or $\mathbf M$.

\subsection{Nearest-neighbor structure factor $f_1$}

The nearest-neighbor structure factor is defined as
\begin{equation}
f_1(\mathbf k)=\sum_{i=1}^{3}e^{i\mathbf k\cdot\boldsymbol\delta_i},
\end{equation}
with nearest-neighbor vectors
\begin{align}
\boldsymbol\delta_1&=-\frac{\sqrt3}{3}\mathbf e_x, \notag\\
\boldsymbol\delta_2&=\frac{\sqrt3}{6}\mathbf e_x-\frac12\mathbf e_y, \notag\\
\boldsymbol\delta_3&=\frac{\sqrt3}{6}\mathbf e_x+\frac12\mathbf e_y.
\end{align}
Expanding around the valleys $\mathbf K^\xi$ gives
\begin{align}
f_1(\mathbf K^\xi+\mathbf q)
&=
\left(-\frac34+i\frac{\sqrt3}{4}\right)q_x
+
\xi\left(\frac{\sqrt3}{4}+i\frac34\right)q_y
\notag\\[1mm]
&\quad
+
\left(\frac1{16}+i\frac{\sqrt3}{16}\right)q_x^2
+
\xi\left(-\frac{\sqrt3}{8}+i\frac18\right)q_xq_y
-
\left(\frac1{16}+i\frac{\sqrt3}{16}\right)q_y^2
\notag\\[1mm]
&\quad
+
o(|\mathbf q|^2).\\[1mm]
&=
\frac{\sqrt{3}}{2} e^{i\frac{5\pi}{6}} q e^{-i\xi \theta_q}
\quad + \quad \frac18 e^{i\frac{\pi}{3}} q^2 e^{i 2\xi\theta_q}
\quad + \quad o(|\mathbf q|^2).
\label{eq:f1_K_expansion}
\end{align}

Around the $M$ point, one obtains
\begin{align}
f_1(\mathbf M+\mathbf q)
&=
\frac12+i\frac{\sqrt3}{2}
+
\left(-1+i\frac{\sqrt3}{3}\right)q_x
\notag\\[1mm]
&\quad
+
\left(\frac1{24}+i\frac{\sqrt3}{24}\right)q_x^2
-
\left(\frac18+i\frac{\sqrt3}{8}\right)q_y^2
+
o(|\mathbf q|^2).
\label{eq:f1_M_expansion}
\end{align}

\subsection{Second-neighbor structure factor $f_2$}
\label{app:f2}
The second-neighbor structure factor is
\begin{equation}
f_2(\mathbf k)=\sum_{i=1}^{6}e^{i\mathbf k\cdot\boldsymbol\delta_i'},
\end{equation}
where
\begin{align}
\boldsymbol\delta'_1&=\mathbf a_1, &
\boldsymbol\delta'_2&=\mathbf a_2, &
\boldsymbol\delta'_3&=\mathbf a_2-\mathbf a_1, \notag\\
\boldsymbol\delta'_4&=-\mathbf a_1, &
\boldsymbol\delta'_5&=-\mathbf a_2, &
\boldsymbol\delta'_6&=\mathbf a_1-\mathbf a_2.
\end{align}
Its expansion around $\mathbf K^\xi$ reads
\begin{align}
f_2(\mathbf K^\xi+\mathbf q)
&=
-3
+
\frac34\left(q_x^2+q_y^2\right)
+
o(|\mathbf q|^2),
\label{eq:f2_K_expansion}
\end{align}
while around $\mathbf M$ one finds
\begin{align}
f_2(\mathbf M+\mathbf q)
&=
-2
+
\frac32 q_x^2
-
\frac12 q_y^2
+
o(|\mathbf q|^2).
\label{eq:f2_M_expansion}
\end{align}

\subsection{Third-neighbor structure factor $f_3$}

The third-neighbor structure factor is defined by
\begin{equation}
f_3(\mathbf k)=\sum_{i=1}^{3}e^{i\mathbf k\cdot\boldsymbol\delta_i''},
\end{equation}
with
\begin{align}
\boldsymbol\delta''_1&=\mathbf a_1+\mathbf a_2+\boldsymbol\delta_1, \notag\\
\boldsymbol\delta''_2&=\mathbf a_2-\mathbf a_1+\boldsymbol\delta_1, \notag\\
\boldsymbol\delta''_3&=\mathbf a_1-\mathbf a_2+\boldsymbol\delta_1.
\end{align}
The valley expansion is
\begin{align}
f_3(\mathbf K^\xi+\mathbf q)
&=
\left(\frac32-i\frac{\sqrt3}{2}\right)q_x
-
\xi\left(\frac{\sqrt3}{2}+i\frac32\right)q_y
\notag\\[1mm]
&\quad
+
\left(\frac14+i\frac{\sqrt3}{4}\right)q_x^2
-
\xi\left(\frac{\sqrt3}{2}-i\frac12\right)q_xq_y
-
\left(\frac14+i\frac{\sqrt3}{4}\right)q_y^2
\notag\\[1mm]
&\quad
+
o(|\mathbf q|^2).
\label{eq:f3_K_expansion}
\end{align}
At the $M$ point, the expansion becomes
\begin{align}
f_3(\mathbf M+\mathbf q)
&=
-\frac32-i\frac{3\sqrt3}{2}
+
\left(\frac12+i\frac{\sqrt3}{2}\right)
\left(q_x^2+q_y^2\right)
+
o(|\mathbf q|^2).
\label{eq:f3_M_expansion}
\end{align}

\subsection{Spin-orbit coupling structure factor $F$}
\label{app:socF}
The spin-orbit structure factor associated with second-neighbor Kane--Mele-type hoppings is
\begin{equation}
F(\mathbf k)
=
-2\left[
\sin(\mathbf k\cdot\boldsymbol\delta'_1)
-
\sin(\mathbf k\cdot\boldsymbol\delta'_2)
+
\sin(\mathbf k\cdot\boldsymbol\delta'_3)
\right].
\end{equation}
Its low-energy expansion around the valleys is
\begin{align}
F(\mathbf K^\xi+\mathbf q)
&=
3\sqrt3\,\xi
-
\xi\frac{3\sqrt3}{4}
\left(q_x^2+q_y^2\right)
+
o(|\mathbf q|^2),
\label{eq:F_K_expansion}
\end{align}
whereas around the $M$ point one obtains
\begin{align}
F(\mathbf M+\mathbf q)
&=
-4q_y
+
o(|\mathbf q|^2).
\label{eq:F_M_expansion}
\end{align}

\section{Green's functions}
\label{sec:Greens}

\subsection{General formalism}

The impurity-induced bound states discussed in the main text are naturally described within the Green's function formalism. In this section, we derive the real-space Green's function associated with the generic continuum Hamiltonian introduced in Sec.~\ref{ssec:General}. The resulting expressions apply to most models considered in this work and constitute the starting point for the calculation of the impurity-induced local density of states.

The real-space Green's function $G_0$ is obtained from its momentum-space representation, $\widetilde G_0$ through a Fourier transform,
\begin{align}
G_0(\mathbf r,0;\omega)
&=
\langle\mathbf r|G_0(\omega)|0\rangle
\nonumber\\
&=
\sum_{\mathbf k,\mathbf k'}
\langle\mathbf r|\mathbf k\rangle
\langle\mathbf k|\widetilde G_0(\omega)|\mathbf k'\rangle
\langle\mathbf k'|0\rangle.
\end{align}

Because the pristine crystal is translationally invariant, the Green's function is diagonal in momentum space,
\begin{equation}
\langle\mathbf k|\widetilde G_0(\omega)|\mathbf k'\rangle
=
\delta_{\mathbf k,\mathbf k'}
\,\widetilde G_0(\mathbf k;\omega),
\end{equation}
so that
\begin{align}
G_0(\mathbf r,0;\omega)
&=
\frac1N
\sum_{\mathbf k}
e^{-i\mathbf k\cdot\mathbf r}
\widetilde G_0(\mathbf k;\omega)
\nonumber\\
&\xrightarrow[N\rightarrow\infty]{}
\int_{\mathrm{BZ}}
\frac{d^2k}{A_{\mathrm{BZ}}}
e^{-i\mathbf k\cdot\mathbf r}
\widetilde G_0(\mathbf k;\omega),
\label{eq:FourierGreen}
\end{align}
where $A_{\mathrm{BZ}}$ denotes the area of the Brillouin zone.

The momentum-space Green's function is defined as
\begin{equation}
\widetilde G_0(\mathbf k;\omega)
=
\left[
\omega-H_0(\mathbf k)
\right]^{-1}.
\end{equation}

For the generic two-band Hamiltonian
\begin{equation}
H_0(\mathbf k)
=
h_0(\mathbf k)\mathbb 1
+
\mathbf h(\mathbf k)\cdot\boldsymbol\sigma,
\end{equation}
its inverse can be evaluated analytically, yielding
\begin{equation}
\widetilde G_0(\mathbf k;\omega)
=
\frac{
(\omega-h_0)\mathbb 1
+
\mathbf h\cdot\boldsymbol\sigma
}
{
(\omega-h_0)^2
-
|\mathbf h|^2
}.
\label{eq:genericGF}
\end{equation}

Near an isolated valley $\mathbf Q$, the continuum Hamiltonian introduced in Sec.~\ref{ssec:General} takes the form
\begin{equation}
H_0(\mathbf Q+\mathbf q)
=
\begin{pmatrix}
h_0(\mathbf Q) + \Delta_{\mathbf Q}
&
\lambda^*
|\mathbf q|^n
e^{-iW\theta_{\mathbf q}}
\\
\lambda
|\mathbf q|^n
e^{iW\theta_{\mathbf q}}
&
h_0(\mathbf Q) - \Delta_{\mathbf Q}
\end{pmatrix},
\end{equation}

where $\Delta_{\mathbf Q}$ is the local gap, $\lambda$ is the complex velocity (or inverse effective mass), $n$ is the order of the band crossing, and $W$ denotes the winding number of the off-diagonal phase. We also introduced the shifted energy
\begin{equation}
\widetilde\omega=\omega-h_0(\mathbf Q),
\end{equation}
and
\begin{equation}
\Omega^2
=
\Delta_{\mathbf Q}^2
-
\widetilde\omega^{\,2}.
\end{equation}

the continuum Green's function becomes
\begin{equation}
\widetilde G_0(\mathbf Q+\mathbf q;\omega)
=
-
\frac{
1
}
{
\Omega^2
+
|\lambda|^2
|\mathbf q|^{2n}
}
\begin{pmatrix}
\widetilde\omega+\Delta_{\mathbf Q}
&
\lambda^*
|\mathbf q|^n
e^{-iW\theta_{\mathbf q}}
\\
\lambda
|\mathbf q|^n
e^{iW\theta_{\mathbf q}}
&
\widetilde\omega-\Delta_{\mathbf Q}
\end{pmatrix}.
\label{eq:continuumGF}
\end{equation}

Equation~(\ref{eq:continuumGF}), together with the Fourier representation (\ref{eq:FourierGreen}), forms the starting point for the evaluation of the real-space Green's function.

\subsection{Real-space continuum Green's function}
\label{ssec:real_space_green}

We now evaluate the real-space Green's function obtained by inserting the continuum propagator of Eq.~\eqref{eq:continuumGF} into the Fourier representation Eq.~\eqref{eq:FourierGreen}. Keeping only the contributions from the low-energy valleys $\mathbf Q$, one finds
\begin{equation}
G_0(\mathbf r,0;\omega)
\simeq
\sum_{\mathbf Q}
e^{-i\mathbf Q\cdot\mathbf r}
\int
\frac{d^2q}{A_{\mathrm{BZ}}}
e^{-i\mathbf q\cdot\mathbf r}
\widetilde G_0(\mathbf Q+\mathbf q;\omega).
\label{eq:valley_decomposition_GF}
\end{equation}

The radial integrals are controlled by the denominator
\begin{equation}
\Omega^2+|\lambda|^2q^{2n}.
\end{equation}
It is convenient to introduce the dimensionless momentum
\begin{equation}
p=\left(\frac{|\lambda|}{\Omega}\right)^{1/n}q,
\end{equation}
and the corresponding dimensionless distance
\begin{equation}
\widetilde r=
\left(\frac{\Omega}{|\lambda|}\right)^{1/n}r.
\end{equation}
The poles of the radial integrals are determined by
\begin{equation}
1+p^{2n}=0.
\end{equation}
We denote by $\zeta_j$, $j=0,\ldots,n-1$, the poles lying in the upper half of the complex plane. With this convention, the real-space Green's function can be expressed in terms of modified Bessel functions.

For the diagonal components, the angular integration is trivial and gives
\begin{align}
G_0(\mathbf r,0;\omega)_{AA}
&\simeq
\frac{2\pi}{A_{\mathrm{BZ}}}
\sum_{\mathbf Q}
e^{-i\mathbf Q\cdot\mathbf r}
\frac{\widetilde\omega+\Delta_{\mathbf Q}}
{n|\lambda|^{2/n}\Omega^{2(1-1/n)}}
\sum_{j=0}^{n-1}
\zeta_j^2
K_0(-i\zeta_j\widetilde r),
\label{eq:G_AA_real_space}
\\[1mm]
G_0(\mathbf r,0;\omega)_{BB}
&\simeq
\frac{2\pi}{A_{\mathrm{BZ}}}
\sum_{\mathbf Q}
e^{-i\mathbf Q\cdot\mathbf r}
\frac{\widetilde\omega-\Delta_{\mathbf Q}}
{n|\lambda|^{2/n}\Omega^{2(1-1/n)}}
\sum_{j=0}^{n-1}
\zeta_j^2
K_0(-i\zeta_j\widetilde r).
\label{eq:G_BB_real_space}
\end{align}

The off-diagonal components contain the phase winding of the continuum Hamiltonian. For $n\equiv W\pmod{2}$, the $BA$ component reads
\begin{align}
G_0(\mathbf r,0;\omega)_{BA}
&\simeq
\frac{2\pi}{A_{\mathrm{BZ}}}
\sum_{\mathbf Q}
e^{-i\mathbf Q\cdot\mathbf r}
\frac{(-1)^W\lambda\,e^{iW\theta_{\mathbf r}}}
{n|\lambda|^{1+2/n}\Omega^{1-2/n}}
\sum_{j=0}^{n-1}
\zeta_j^{n+2}
K_W(-i\zeta_j\widetilde r),
\label{eq:G_BA_real_space}
\end{align}
while the $AB$ component is obtained by complex conjugation of the winding structure,
\begin{align}
G_0(\mathbf r,0;\omega)_{AB}
&\simeq
\frac{2\pi}{A_{\mathrm{BZ}}}
\sum_{\mathbf Q}
e^{-i\mathbf Q\cdot\mathbf r}
\frac{(-1)^W\lambda^*\,e^{-iW\theta_{\mathbf r}}}
{n|\lambda|^{1+2/n}\Omega^{1-2/n}}
\sum_{j=0}^{n-1}
\zeta_j^{*(n+2)}
K_W(i\zeta_j^*\widetilde r).
\label{eq:G_AB_real_space}
\end{align}

Equivalently, the result can be written in matrix form as
\begin{align}
G_0(\mathbf r,0;\omega)
&\simeq
\frac{2\pi}{A_{\mathrm{BZ}}}
\sum_{\mathbf Q}
\frac{e^{-i\mathbf Q\cdot\mathbf r}}
{n|\lambda|^{2/n}\Omega^{2(1-1/n)}}
\sum_{j=0}^{n-1}
\begin{pmatrix}
\mathcal G_{AA}^{(j)}(\mathbf r,\omega)
&
\mathcal G_{AB}^{(j)}(\mathbf r,\omega)
\\
\mathcal G_{BA}^{(j)}(\mathbf r,\omega)
&
\mathcal G_{BB}^{(j)}(\mathbf r,\omega)
\end{pmatrix},
\end{align}
with
\begin{align}
\mathcal G_{AA}^{(j)}
&=
\quad (\widetilde\omega+\Delta_{\mathbf Q}) \quad
\zeta_j^2
K_0(-i\zeta_j\widetilde r),
\\
\mathcal G_{BB}^{(j)}
&=
\quad (\widetilde\omega-\Delta_{\mathbf Q}) \quad
\zeta_j^2
K_0(-i\zeta_j\widetilde r),
\\
\mathcal G_{BA}^{(j)}
&=
\frac{(-1)^W\lambda}
{|\lambda|} e^{iW\theta_{\mathbf r}} \quad \Omega \quad 
\zeta_j^{(n+2)}
K_{|W|}(-i\zeta_j\widetilde r),
\\
\mathcal G_{AB}^{(j)}
&=
\frac{(-1)^W\lambda^*}
{|\lambda|} e^{-iW\theta_{\mathbf r}} \quad \Omega \quad 
\zeta_j^{*(n+2)}
K_{|W|}(+i\zeta^*_j\widetilde r).
\end{align}

These expressions make explicit the two ingredients relevant for the impurity-induced LDOS: the exponential spatial decay governed by the modified Bessel functions and the phase winding carried by the off-diagonal components through the factor $e^{\pm iW\theta_{\mathbf r}}$.

\subsection{Impurity-induced local density of states}

We now derive the impurity-induced contribution to the local density of states (LDOS) associated with a bound state. Within the $T$-matrix formalism, the correction to the Green's function produced by a localized impurity of strength $V_0$ acting on the $A$ sublattice is
\begin{equation}
\delta G(\mathbf r,\mathbf r;\omega)
=
G_0(\mathbf r,0;\omega)\,
T(\omega)\,
G_0(0,\mathbf r;\omega),
\end{equation}
where the $T$ matrix is
\begin{equation}
T(\omega)
=
\frac{V_0}
{1-V_0G_0(0,0;\omega)_{AA}}.
\end{equation}
The impurity-induced LDOS then follows from
\begin{equation}
\delta\rho(\mathbf r,\omega)
=
-\frac1\pi
\Im
\sum_\alpha
\delta G(\mathbf r,\mathbf r;\omega)_{\alpha\alpha},
\end{equation}
or equivalently,
\begin{align}
\delta\rho(\mathbf r,\omega)
=
-\frac1\pi
\sum_\alpha
\Im
\left[
\frac{
G_0(\mathbf r,0;\omega)_{\alpha A}
V_0
G_0(0,\mathbf r;\omega)_{A\alpha}
}
{
1-V_0G_0(0,0;\omega)_{AA}
}
\right].
\label{eq:LDOS_start}
\end{align}

Inside the bulk gap, the local Green's function can be written as
\begin{equation}
G_0(0,0;\omega)_{AA}
=
R_0(\omega)+iI_0(\omega),
\end{equation}
while the spectral density vanishes and therefore $I_0(\omega)\xrightarrow[\eta\rightarrow0^+]{}0$ vanishes in the limit $\eta\rightarrow0^+$. Consequently, the denominator of Eq.~\eqref{eq:LDOS_start} develops a simple pole whenever
\begin{equation}
1-V_0R_0(\omega_b)=0,
\label{eq:bound_state_condition_appendix}
\end{equation}
which is precisely the bound-state condition derived in the main text.

Near the bound-state energy $\omega_b$, we expand
\begin{equation}
1-V_0R_0(\omega)
\simeq
-V_0R_0'(\omega_b)(\omega-\omega_b),
\end{equation}
so that the imaginary part of the $T$ matrix becomes
\begin{align}
-\frac1\pi\Im T(\omega)
&=
-\frac1\pi
\Im
\left[
\frac{V_0}
{-V_0R_0'(\omega_b)(\omega-\omega_b)-iV_0I_0}
\right]
\nonumber\\
&\xrightarrow[\eta\rightarrow0^+]{}
\frac{\delta(\omega-\omega_b)}
{|R_0'(\omega_b)|},
\end{align}
where we used the distributional identity
\begin{equation}
\lim_{\eta\rightarrow0^+}
\frac1\pi
\frac{\eta}
{x^2+\eta^2}
=
\delta(x).
\end{equation}

The impurity-induced LDOS therefore assumes the compact form
\begin{equation}
\delta\rho(\mathbf r,\omega)
=
\frac{\delta(\omega-\omega_b)}
{|R_0'(\omega_b)|}
\sum_\alpha
\Re
\!\left[
G_0(\mathbf r,0;\omega_b)_{\alpha A}
G_0(0,\mathbf r;\omega_b)_{A\alpha}
\right].
\label{eq:LDOS_final}
\end{equation}

Equation~\eqref{eq:LDOS_final} shows that the spatial structure of the bound-state LDOS is entirely determined by the pristine Green's function derived in the previous subsection, while the spectral dependence reduces to a delta peak at the impurity-induced bound-state energy.

\section{Spin--orbit coupling conventions}
\label{app:SOC_conventions}

The spin--orbit sector of the Hamiltonian consists of two second-nearest-neighbor hopping terms on the honeycomb lattice: the intrinsic Kane--Mele (KM) coupling and the valley--Zeeman (VZ) coupling. Both correspond to purely imaginary hoppings between sites belonging to the same sublattice. Their relative signs depend on the orientation of the hopping path around the hexagonal plaquettes.

\begin{figure}[h!]
\centering
\begin{minipage}{0.95\textwidth}
\centering

\begin{tikzpicture}
[
    scale=1.1,
    hex/.style={thick},
    hop/.style={very thick},
    arr/.style={-{Latex[length=2mm]}, thick},
    lab/.style={font=\scriptsize},
    vec/.style={font=\scriptsize}
]

\newcommand{\hexagon}[3]{%
    \coordinate (O) at #1;
    \foreach \i in {0,...,5}{
        \coordinate (P\i) at ($(O)+({60 + 60*\i}:0.75)$);
    }
    \draw[hex] (P0)--(P1)--(P2)--(P3)--(P4)--(P5)--cycle;

    \foreach \i in {0,2,4} \fill (P\i) circle (1.2pt);
    \foreach \i in {1,3,5} \fill (P\i) circle (1.2pt);

    \ifnum#3=0
        \draw[hop,#2] (P0)--(P2)--(P4)--cycle;
    \else
        \draw[hop,#2] (P1)--(P3)--(P5)--cycle;
    \fi
}

\newcommand{\signvec}[3]{%
    \node[vec,anchor=west] at #1 {$
    \lambda^{#2}_{#3} =
    \begin{pmatrix}
    \phantom{-}1\\[-1mm]
    -1\\[-1mm]
    \phantom{-}1
    \end{pmatrix}$};
}

\newcommand{\signvecminus}[3]{%
    \node[vec,anchor=west] at #1 {$
    \lambda^{#2}_{#3} =
    \begin{pmatrix}
    -1\\[-1mm]
    \phantom{-}1\\[-1mm]
    -1
    \end{pmatrix}$};
}

\definecolor{Acolor}{RGB}{230,140,20}      
\definecolor{Bcolor}{RGB}{0,140,140}       
\definecolor{PlusColor}{RGB}{0,120,0}      
\definecolor{MinusColor}{RGB}{170,30,30}   
\definecolor{ArrowColor}{gray}{0.55}

\node at (0,1.25) {$K\!M$};
\node at (4.0,1.25) {$V\!Z$};

\hexagon{(0,0)}{red}{0}
\node[lab] at (-1.05,-0.0) {$A$};
\node[lab] at (-0.75,-0.65) {$B$};
\draw[arr,ArrowColor,-](-0.18,-0.15)
arc[start angle=220,end angle=520,radius=0.23];
\node[ArrowColor,fill=ArrowColor,regular polygon,regular polygon sides=3,
  inner sep=0pt,
  minimum size=5pt,
  rotate=40
] at (-0.22,0.06) {};
\node[PlusColor,font=\scriptsize] at (-0.25,0.45) {$-$};
\node[orange,font=\scriptsize] at (0.54,-0.0) {$+$};
\node[orange,font=\scriptsize] at (-0.25,-0.45) {$+$};
\signvec{(1.0,0.05)}{A}{KM}

\hexagon{(0,-2.0)}{blue}{1}
\node[lab] at (-1.05,-2) {$A$};
\node[lab] at (-0.75,-2.65) {$B$};
\draw[arr,ArrowColor,-](-0.18,-2.15)
arc[start angle=220,end angle=520,radius=0.23];
\node[ArrowColor,fill=ArrowColor,regular polygon,regular polygon sides=3,
  inner sep=0pt,
  minimum size=5pt,
  rotate=40
] at (-0.22,-1.94) {};
\node[PlusColor,font=\scriptsize] at (0.25,-1.55) {$-$};
\node[PlusColor,font=\scriptsize] at (-0.54,-2.0) {$-$};
\node[orange,font=\scriptsize] at (0.25,-2.45) {$+$};
\signvecminus{(1.0,-1.95)}{B}{KM}

\hexagon{(4,0)}{red}{0}
\draw[arr,ArrowColor,-] (3.82,-0.15)
arc[start angle=220,end angle=520,radius=0.23];
\node[ArrowColor,fill=ArrowColor,regular polygon,regular polygon sides=3,
  inner sep=0pt,
  minimum size=5pt,
  rotate=40
] at (3.78,0.06) {};
\node[PlusColor,font=\scriptsize] at (3.75,0.45) {$-$};
\node[orange,font=\scriptsize] at (4.54,-0.0) {$+$};
\node[orange,font=\scriptsize] at (3.75,-0.45) {$+$};
\signvec{(5.0,0.05)}{A}{VZ}

\hexagon{(4,-2.0)}{blue}{1}
\draw[arr,ArrowColor,-] (4.18,-1.85) 
arc[start angle=40,end angle=-260,radius=0.23];
\node[ArrowColor,fill=ArrowColor,regular polygon,regular polygon sides=3,
  inner sep=0pt,
  minimum size=5pt,
  rotate=-90
] at (3.97,-1.78) {};
\node[orange,font=\scriptsize] at (4.25,-1.55) {$+$};
\node[orange,font=\scriptsize] at (3.46,-2.0) {$+$};
\node[PlusColor,font=\scriptsize] at (4.25,-2.45) {$-$};
\signvec{(5.0,-1.95)}{B}{VZ}
\end{tikzpicture}
\caption{Sign conventions adopted for the intrinsic Kane--Mele (left) and valley--Zeeman (right) second-nearest-neighbor hoppings.
Red and blue triangles correspond to the two inequivalent plaquette orientations. The indicated signs define the imaginary hopping amplitudes entering the lattice Hamiltonian and are used consistently throughout this work.}
\label{fig:KM_VZ_conventions}
\end{minipage}
\end{figure}
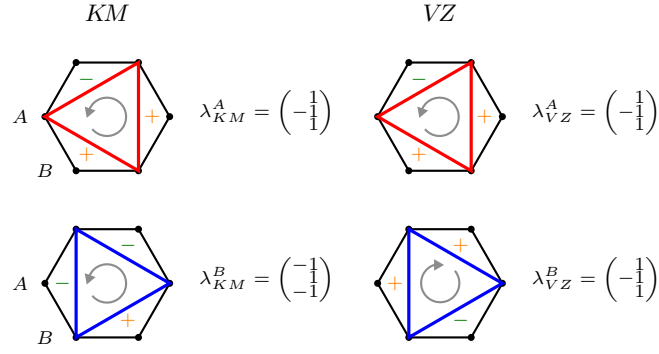

Throughout this work, we adopt the convention illustrated in Fig.~\ref{fig:KM_VZ_conventions}. The Kane--Mele hopping changes sign between the two sublattices. Conversely, the valley--Zeeman hopping has the same sign on both sublattices.

With these conventions, the spin--orbit Hamiltonian introduced in Sec.~\ref{ssec:SOC} reads
\begin{align}
H_{\mathrm{KM}}(\mathbf k)
&=
t_{\mathrm{KM}}
F(\mathbf k)
s_z\sigma_z,
\\
H_{\mathrm{VZ}}(\mathbf k)
&=
t_{\mathrm{VZ}}
F(\mathbf k)
s_z\sigma_0,
\end{align}
where the $s$ matrices are in spin-basis, the $\sigma$ matrices are in sublattice-basis, and the structure factor
\begin{equation}
F(\mathbf k)
=
-2
\left[
\sin(\mathbf k\!\cdot\!\boldsymbol\delta'_1)
-
\sin(\mathbf k\!\cdot\!\boldsymbol\delta'_2)
+
\sin(\mathbf k\!\cdot\!\boldsymbol\delta'_3)
\right]
\end{equation}
was introduced in Appendix~\ref{sec:structure_factor_expansions}.

\section{Filtering}
\label{sec:Filtering}

The extraction of the winding from the LDOS is not entirely independent of the filtering procedure. In particular, the width $\sigma$ of the Gaussian filter applied in momentum space controls the compromise between spatial and momentum resolution. A narrower filter isolates more precisely the QPI feature around $\pm\Delta \mathbf{K}$, but broadens the reconstructed signal in real space and may wash out short-distance features. Conversely, a wider filter preserves the behavior close to the impurity at the expense of a poorer separation of the momentum-space contributions. This sensitivity is a direct consequence of the Fourier uncertainty principle.

\begin{figure}[h!]
\includegraphics[width=0.46\textwidth]{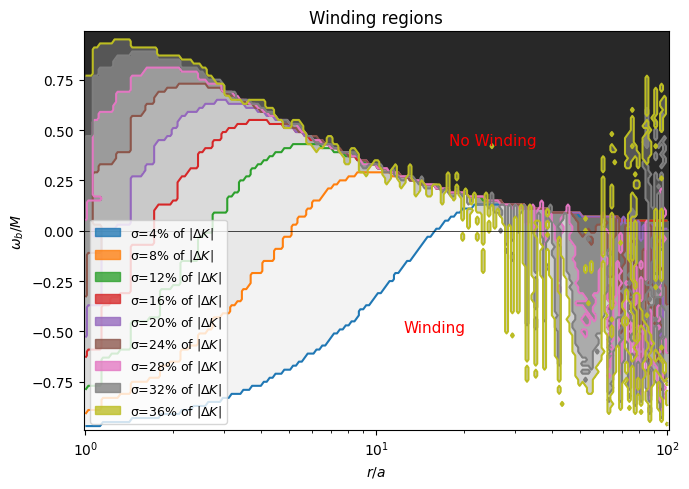}
\caption{Winding regions as a function of the filtering width $\sigma$. Each curve corresponds to a different value of $\sigma$. The region below a given curve exhibits a non-trivial winding, whereas no winding is observed above it.}
\label{fig:per_sigma}
\end{figure}

Figure~\ref{fig:per_sigma} illustrates this dependence by showing the regions of parameter space where a non-trivial winding is observed for different values of $\sigma$. While the quantitative boundary depends on the filter width, the overall behavior remains robust: increasing $\sigma$ extends the region where the winding can be reliably detected, particularly in the vicinity of the impurity on the scale of the lattice constant.

\section{Semi-Dirac model details}
\label{app:Hybrid}

\subsection{Semi-Dirac engineering in the NNNN lattice model}
\label{engineering}
We consider the two-band Hamiltonian defined by
$$
H(\mathbf{k})=
\begin{pmatrix}
M+t_2 f_2(\mathbf{k}) & t_1 f_1(\mathbf{k})+t_3 f_3(\mathbf{k}) \\
\big(t_1 f_1(\mathbf{k})+t_3 f_3(\mathbf{k})\big)^* & -\big(M+t_2 f_2(\mathbf{k})\big)
\end{pmatrix},
$$
with dispersion

$$\varepsilon(\mathbf{k})=\sqrt{|t_1 f_1(\mathbf{k})+t_3 f_3(\mathbf{k})|^2+\big(M+t_2 f_2(\mathbf{k})\big)^2}.$$

Our goal is to realize a semi-Dirac dispersion at the $\mathbf{M}$-point, characterized by linear dispersion along one direction and quadratic dispersion along the orthogonal direction, while allowing for a small gap opening.

\subsection{Parameter choice}
\label{Parameter choice}

Expanding around ($\mathbf{k}=\mathbf{M}+\mathbf{q}$), the constant terms are

$$f_1(\mathbf{M})=e^{i\pi/3},\quad f_3(\mathbf{M})=-3e^{i\pi/3},\quad f_2(\mathbf{M})=-2.$$

A band touching at ($\mathbf{M}$) requires cancellation of the constant contributions. This leads to the semi-Dirac tuning

$$t_1=3t_3,\qquad M=2t_2.$$

To allow for a small gap at ($\mathbf{M}$), we introduce a detuning

$$M=2t_2+\delta,\qquad |\delta|\ll t_1.$$

\subsection{Expansion of the diagonal and off-diagonal terms}
\label{Expansion off-diagonal}

Using the Taylor expansions of ($f_1$) and ($f_3$) App.~\ref{sec:structure_factor_expansions}, we obtain at the semi-Dirac tuning ($t_1=3t_3\equiv 3t$), up to quadratic order:
$$H_0(\mathbf{M} + \mathbf{q})_{AA} =\delta - \frac{t_2}{2}q_y^2 + o(q_x,q_y^2).$$
At the critical point ($\delta=0$), the leading contribution is purely quadratic. Similarly,

$$e^{-i\pi/3} H_0(\mathbf{M} + \mathbf{q})_{BA} = i2\sqrt{3}t q_x -\frac{t}{4} q_y^2 +  o(q_x,q_y^2).$$

Thus, the leading contribution is linear in ($q_x$), while no linear term appears in ($q_y$).

\subsection{Resulting dispersion}
\label{Expansion dispersion}

If $\delta$ is small enough, the low-energy dispersion follows as:

$$\varepsilon^2(\vecM+\mathbf{q}) = \delta^2 + 12 t^2 q_x^2 + \left(\frac{t_2^2}{4}+\frac{t^2}{16}\right) q_y^4 + o(q_x^2,q_y^4),$$
which yields
  $$
  \varepsilon(\vecM+\mathbf{q})
  \sim |\delta| + |v_x| |q_x| + \beta q_y^2
  $$
  in the right range of $\delta/t \ll q \ll \Delta \mathbf{K}$ with
  $$
  v_x = e^{i\pi/3}2\sqrt{3}t,\qquad
  \beta = \sqrt{\frac{t_2^2}{4}+\frac{t^2}{16}}.
  $$
This demonstrates the semi-Dirac character: linear dispersion along ($q_x$) and quadratic dispersion along ($q_y$).

\subsection{Berry curvature}
\label{Berry Hybrid}

Near the $M$ point, the Berry curvature, in this continuum limit, follows as:
$$
\Omega_\pm(\mathbf q)
=
\pm
\frac{\sqrt{3}\,\delta\,t^2\,q_y}{2
\left[
\left(
\delta+\frac{t_2}{2}(3q_x^2-q_y^2)
\right)^2
+
12t^2 q_x^2
+
\frac{t^2}{16}(5q_x^2+q_y^2)^2
\right]^{3/2}}.
$$

\end{widetext}


\end{document}